\pdfoutput=1
\documentclass[usenatbib,usedAMS]{mnras}

\usepackage{natbib}
\usepackage{amsmath}
\usepackage{graphicx}
\usepackage{color}
\usepackage{mathrsfs}
\usepackage{epsfig}
\usepackage{comment}
\usepackage{ulem}

\newcommand{\msun}{{\rm M}_\odot}
\newcommand{\rsun}{{\rm R}_\odot}
\newcommand{\zsun}{Z_\odot}
\newcommand{\lsun}{{\rm L}_\odot}
\newcommand{\cc}{{\rm cm}^{-3}}

\newcommand{\msunyr}{{\rm M}_\odot~{\rm yr}^{-1}}

\newcommand{\mpc}{{\rm Mpc}}
\newcommand{\gpc}{{\rm Gpc}}

\newcommand{\yr}{\rm yr}

\newcommand{\K}{{\rm K}}
\newcommand{\beq}{\begin{equation}}
\newcommand{\eeq}{\end{equation}}

\defcitealias{K14}{K14}
\defcitealias{K16}{K16}
\defcitealias{Visbal_2015}{VHB15}

\title[PopIII binary BH formation]
{Formation pathway of Population III coalescing binary black holes
through stable mass transfer}

\author[]{Kohei Inayoshi$^{1}$\thanks{E-mail: inayoshi@astro.columbia.edu (KI)}
\thanks{Simons Society of Fellows, Junior Fellow.},
{Ryosuke Hirai}$^2$, {Tomoya Kinugawa}$^3$ and Kenta Hotokezaka$^{4}$
\\
$^{1}$Department of Astronomy, Columbia University, 550 W. 120th Street, New York, NY 10027, USA\\
$^{2}$Advanced Research Institute for Science and Engineering, Waseda University, 3-4-1, Okubo, Shinjuku, Tokyo 169-8555, Japan\\
$^{3}$Institute for Cosmic Ray Research, The University of Tokyo, Chiba 277-8582, Japan\\
$^{4}$Center for Computational Astrophysics, 162 5th Ave, New York, NY, 10010, USA
}

\begin{document}
\maketitle
\label{firstpage}

\begin{abstract}
We study the formation of stellar mass binary black holes (BBHs) originating from Population III (PopIII) stars,
performing stellar evolution simulations for PopIII binaries with {\tt MESA}.
We find that a significant fraction of PopIII binaries form massive BBHs through stable mass transfer 
between two stars in a binary, without experiencing common envelope phases.
We investigate necessary conditions required for PopIII binaries to form coalescing BBHs
with a semi-analytical model calibrated by the stellar evolution simulations.
The BBH formation efficiency is estimated
for two different initial conditions for PopIII binaries with large and small separations, respectively.
Consequently, in both models, $\sim 10\%$ of the total PopIII binaries form BBHs only through stable mass transfer
and $\sim 10\%$ of these BBHs merge due to gravitational wave emission within the Hubble time.
Furthermore, the chirp mass of merging BBHs has a flat distribution over $15\la M_{\rm chirp}/\msun \la 35$.
This formation pathway of PopIII BBHs is presumably robust because stable mass transfer is less uncertain 
than common envelope evolution, which is the main formation channel for Population II BBHs.
We also test the hypothesis that the BBH mergers detected
by LIGO originate from PopIII stars using the total number of PopIII stars formed
in the early universe as inferred from the optical depth measured by Planck.
We conclude that the PopIII BBH formation scenario can explain
the mass-weighted merger rate of the LIGO's O1 events
with the maximal PopIII formation efficiency inferred from the Planck measurement,
even without BBHs formed by unstable mass transfer or common envelope phases.
\end{abstract}



\section{Introduction}
\label{sec:intro}

Advanced LIGO (AdLIGO) has detected sources of gravitational waves (GWs).
The sources, GW150914, GW151226 and LVT151012, are inferred to be merging
binary black holes (BBHs) with masses of ($36.2^{+5.2}_{-3.8}~\msun, 29.1^{+3.7}_{-4.4}~\msun$),
($14.2^{+8.3}_{-3.7}~\msun, 7.5^{+2.3}_{-2.3}~\msun$) and 
($23^{+18}_{-6}~\msun, 13^{+4}_{-5}~\msun$)
\citep{Abbott_PRL_2016,Abbott_PRL_2_2016}.
The origin of such massive and compact BBHs and their formation pathways 
have been proposed 
\citep[][references therein]{Abbott_2016_Astro} through massive binary evolution 
\citep[e.g.][]{Belczynski_2004,Dominik_2012,K14,Belczynski_2016_b}
including rapid rotation and tides \citep{Mandel_2016, 2016A&A...588A..50M},
and/or stellar dynamics in a dense cluster
\cite[e.g.][]{PortegiesZwart_2000,OLeary_2016,Rodriguez_2016,Mapelli_2016}.

In the isolated-binary scenario, metal-poor stars are generically required 
to form massive BHs because of 
inefficient stellar winds and smaller stellar radii.
Many authors have investigated formation channels of BBHs via Population II (hereafter, PopII) stars 
with $Z\la 0.1~\zsun$ \citep[e.g.][]{Dominik_2012,Belczynski_2016_b} and
Population III (hereafter PopIII) with $Z\simeq 0$ stars \citep[e.g.][]{K14,K16,2016arXiv161201524B}.
The initial mass function (IMF) of PopII stars is expected to be less top-heavy \citep{2005ApJ...626..627O}, 
whereas PopIII stars are thought to be typically as massive as $\sim 10-300~\msun$ \citep[e.g.,][]{2014ApJ...781...60H}
and likely to evolve without losing their masses due to stellar winds \citep[e.g.][]{Baraffe_2001,IHO_2013}.
These scenarios may be distinguished in the future by 
the chirp mass distributions \citep{Nakamura_2016}, 
where $M_{\rm chirp}\equiv (M_1M_2)^{3/5}/(M_1+M_2)^{1/5}$ for a BBH with masses of $M_1$ and $M_2$,
and stochastic GW backgrounds \citep{LIGO_back_2016,Hartwig_2016,
Inayoshi_2016,Dvorkin_2016,Nakazato_2016}.

In PopII BBH formation from isolated binaries, 
a significant fraction of the binaries experience unstable mass transfer (MT) 
during the binary evolution \citep[e.g.][]{Dominik_2012,Belczynski_2016_b}.
The unstable MT makes the orbital separation shrink rapidly, 
which results in a common envelope (CE) phase (\citealt{Paczynski_1976,
Iben_1993,Taam_2000,Ivanova_2013}, references therein), 
where one of the stars plunges
into the bloated envelope of the companion star and spirals inwards, 
losing its orbital energy and angular momentum.
If the stellar envelope is successfully ejected
due to the energy deposit and the spiral-in halts, 
a close binary system would be formed or else these stars would merge. 
To form BBHs which can merge within the Hubble time, PopII binaries should experience 
CE phases because their orbital separations tend to be wider due to mass loss (e.g., stellar winds).
Moreover, since PopII giant stars are likely to have convective stellar envelopes, 
the MT would be unstable.
Because of uncertainties about relevant physical processes, however,
the final outcome of binaries after unstable MT and CE phases is  uncertain
\citep[e.g.,][]{Ivanova_2013}.
In population synthesis models, in fact, merging rates of PopII BBHs 
vary by several orders of magnitude depending on the adopted prescriptions 
about the MT and CE \citep{2008ApJS..174..223B,Dominik_2012,Belczynski_2016_b,
2016MNRAS.462.3302E}\footnote{\cite{Pavlovskii_2016} have pointed 
out that massive stars with convective envelope could be more unlikely to experience unstable 
MT and CE than previously expected in PopII BBH formation \citep[e.g.][]{Belczynski_2016_b},
considering a detailed model \citep{Pavlovskii_2015}.
Although their model also might allow PopIII binaries to form more BBHs even via CE phase,
we do not consider the effect to give a conservative discussion.}.

The CE phase also may play a crucial role for the PopIII binary evolution.  
\cite{2016arXiv161201524B} claimed that most of PopIII massive binaries merge in the CE phase
thereby only a small fraction of them evolve to merging BBHs. 
However, as speculated by \cite{K14}, there are PopIII binaries which can form BBHs 
with a coalescence time less than the Hubble time
{\it through stable MT and without experiencing any CE phases}.
This is because PopIII stars with certain masses evolve to compact blue giants 
with a radiative envelope instead of red giants with a convective envelope
\citep[e.g.,][]{Marigo_2001, Ekstrom_2008}. 
The discrepancies between \cite{K14} and \cite{2016arXiv161201524B}
come from the difference in models for PopIII single stellar evolution 
and their prescription to describe binary evolution during CE phases 
(e.g., stability conditions for MT and merger criteria in CE phases).

In this work, we address the following questions: (i) what conditions are required
for the formation of merging PopIII BBHs to avoid the CE phases and 
(ii) what fraction of PopIII binaries evolve to merging BBHs without the CE phase. 
Then we compare the total number of PopIII BBHs required from LIGO's O1 
merger rate with the total mass of the PopIII stars
in the Universe derived from other astronomical observations.
For this purpose, we study PopIII binary evolution with stellar evolution calculations. 
In this paper, for the first time, we perform stellar evolution calculations with 
a public code \texttt{MESA} \citep[version 8845]{MESA1,MESA2,MESA3}
to follow evolution processes of massive PopIII binaries 
(single stellar evolution, orbital evolution and mass transfer; see \S\ref{MESA})
and formation of BBHs.
Applying the results, we make a semi-analytical model to describe those binary processes 
so that the evolutionary tracks match with the results obtained by the detailed 
calculations of stellar structure.
This method is useful to study statistical properties of forming PopIII BBHs
for a wide range of initial conditions, i.e., the masses and and the orbital separation.

The rest of this paper is organized as follows. 
In \S\ref{sec:method}, we describe the methodology of our stellar evolution calculations (\S\ref{MESA})
and semi-analytical model (\S\ref{sec:semiana}).
In \S\ref{sec:PopIIIBBH_result}, we show the simulation results and discuss the necessary conditions 
for formation of PopIII BBHs only via stable MT.
We also estimate formation efficiency of PopIII BBHs for two scenarios for initial conditions of PopIII binaries.
In \S\ref{sec:BBHeff}, we discuss the possibility that coalescing PopIII BBHs can explain all the LIGO sources,
consistently with observations in the current status.
We give discussions and caveats of our calculations in \S\ref{sec:discussion}
and summarize the main conclusions of this paper in \S\ref{sec:summary}.


\section{Methodology}
\label{sec:method}

\subsection{Stellar evolution calculations}
\label{MESA}

\subsubsection{single star evolution}

We consider the evolution of massive PopIII stars in the mass range of $20\leq M/\msun \leq 60$.
The stellar structure is calculated with a public code {\tt MESA}.
We briefly describe the setup of the calculations but more details are shown in \citet{MESA1,MESA2,MESA3}.
We use the mixing length theory to treat convection, 
with the Schwarzschild criterion and a mixing length parameter $1.6$. 
The effects of mass loss due to stellar winds and pulsations are neglected
because they are unlikely to affect the stellar structure for the case of PopIII stars 
\citep[e.g.][]{Baraffe_2001,IHO_2013}.
We do not follow evolution of a star beyond central carbon (C) ignition
because the C burning phase finishes within a few years, in which the binary properties 
(mass, separation etc) do not change.
Instead, we replace the star with a BH (point gravitational source) of the same mass
when the original stellar mass (or the core mass) is massive enough to form a BH.
We here set the critical mass to $M\geq 28~\msun$ in the zero-age main sequence (ZAMS) phase.
\citep[e.g.,][]{K14}, where the corresponding core masses are 
$M_{\rm He(CO)}\ga 9.3~(7.6)~\msun$.

\begin{figure}
\begin{center}
\includegraphics[width=83mm]{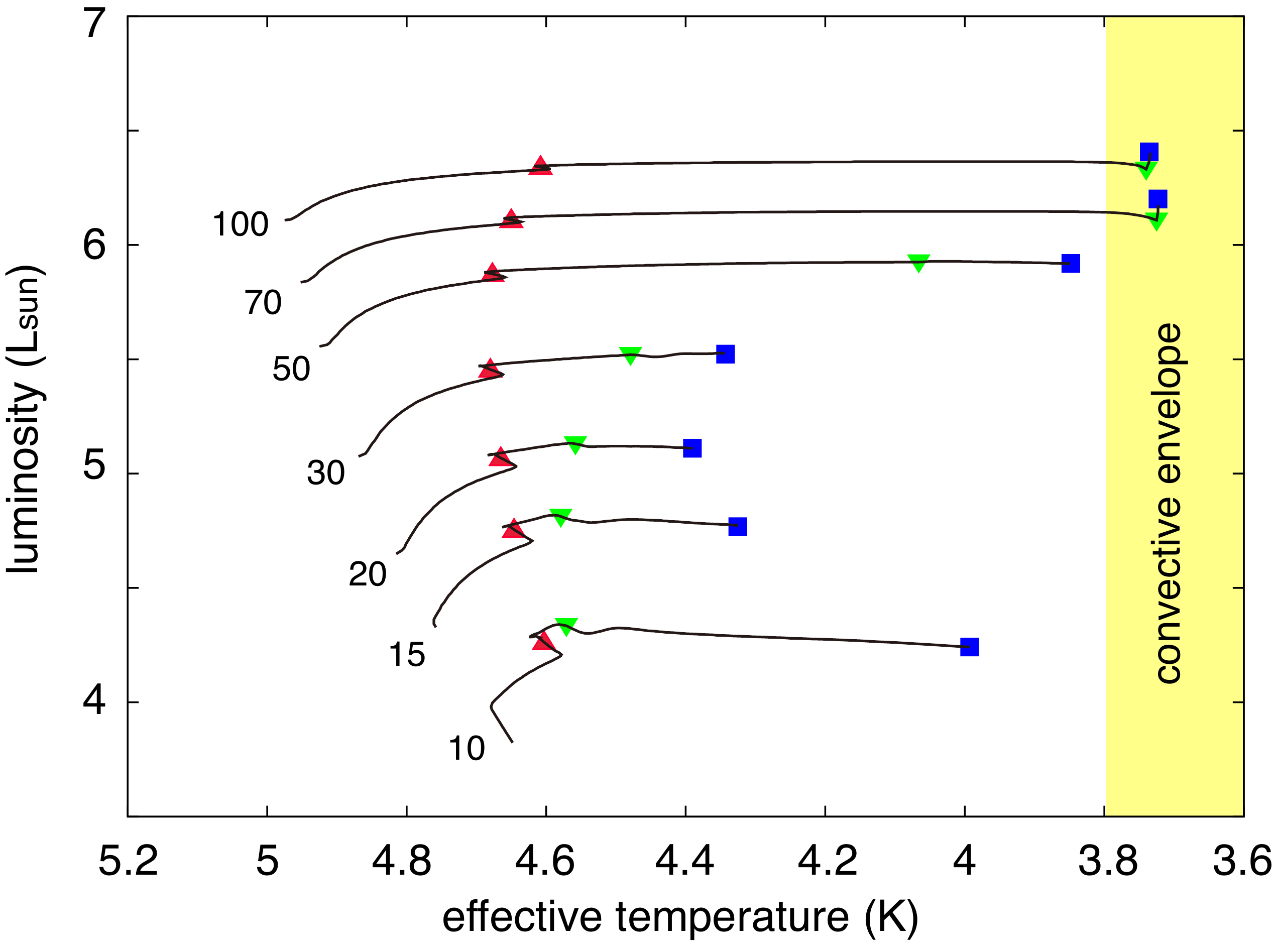}
\caption{The Hertzsprung-Russell (HR) diagram for PopIII stars with a mass range of
$10\leq M/\msun \leq 100$ using the data taken from \citet{Marigo_2001}.
Symbols indicate the epochs corresponding to the beginning of the He-core burning (red), 
the He-shell burning (green) and the C burning (blue), respectively.
}
\label{fig:HR}
\end{center}
\end{figure}

Fig.~\ref{fig:HR} presents the Hertzsprung-Russell (HR) diagram for PopIII stars (solid),
whose data are taken from \citet{Marigo_2001}.
Before including the binary interactions, we check that results of single stellar evolution for 
various masses ($20\leq M\leq 60\msun$)
agree well with those shown in Fig.~\ref{fig:HR}.
Massive PopIII stars are generally more compact and hotter than metal-enriched PopII/I stars.
This is because PopIII stars contract before reaching the ZAMS 
phase so that the central temperature increases to $\sim10^8$ K, 
where helium (He) burning produces a tiny fraction of carbon and thus sufficient 
energy by hydrogen (H) burning due to CN-cycle is generated to support the whole structure \citep[e.g.][]{Omukai_Palla_2003}.
As a result, the effective temperature is $4\times 10^4\la T_{\rm eff}\la 10^5$ K in the ZAMS phase.
For $10\la M/\msun \la 100$, the onset of central He-burning occurs soon 
after hydrogen exhaustion in the core.
The evolution stage proceeds with an almost constant luminosity towards lower effective temperature.
One remarkable feature is that for $10\la M\la 60~\msun$ stars the He-core burning and the subsequent nuclear burning 
terminates before reaching the Hayashi track ($T_{\rm eff}\sim 10^{3.7}$ K), 
where stars develop a deep convective envelope \citep{Hayashi_1961,HHS_1962}.
Detailed calculations of stellar structure by other groups \citep[e.g.,][]{Ekstrom_2008}
have also shown that a single PopIII star with $M\la 60~\msun$ is unlikely to have 
a convective envelope during its lifetime (see their figure 5).
Note that properties of the stellar envelope are crucial to
determine stability of MT (see \S\ref{sec:MT}).

Recently, \cite{2016arXiv161201524B} performed population synthesis calculations of PopIII binaries.
They describe the time evolution of the stellar radius of PopIII stars, by adopting 
their lowest metallicity model ($Z=0.0001$) and assuming that deep convective envelopes 
are developed even with higher values of $T_{\rm eff}$ depending on stellar masses,
e.g., the effective temperatures below which a convective envelope develops are
assumed to be $10^{4.4}$ and $10^{4.3}~\K$ for $M=20$ and $50~\msun$, respectively. 
These correspond to the evolutionary stages when
the core helium burning begins. Under such an assumption,
MT always results in the CE phase and the outcome is a stellar merger.
However, stellar evolution calculations show that PopIII stars do not have 
convective envelopes at $T_{\rm eff}>10^{3.7-3.8}~\K$ \citep[e.g.,][]{Ekstrom_2008,Hosokawa_2012}, 
that is satisfied at any stage of the late-time evolution of PopIII stars with masses $<60M_{\odot}$
as shown in Fig.~\ref{fig:HR}. Therefore we do not
expect that the MT between PopIII stars always leads to the CE phase.

\subsubsection{binary orbital evolution}
\label{sec:orbit}

We consider the orbital evolution of a binary system of
the primary and secondary star with a mass of $M_1$ and $M_2$, respectively.
The time evolution of the orbital separation $a$ is calculated by an ordinary differential equation,
\begin{equation}
\frac{\dot{a}}{a}=\frac{2\dot{J}_{\rm orb}}{J_{\rm orb}}-\frac{2\dot{M}_1}{M_1}
-\frac{2\dot{M}_2}{M_2}+\frac{\dot{M}}{M},
\end{equation}
where $J_{\rm orb}$ is the orbital angular momentum of the binary and 
$M=M_1+M_2$ is the total mass.
As the binary evolves, the mass and the orbital angular momentum will change 
due to mass transfer (MT) between the two stars and emission of GWs after 
they collapse into compact objects.
Note that circular orbits ($e=0$) are assumed for simplicity (but see also \citealt{2016ApJ...825...70D,2016ApJ...825...71D}).

For conservative MT, i.e., $\dot{J}_{\rm orb}=0$ and $\dot{M}=0$ 
($\dot{M}_1=-\dot{M}_2 \neq 0$), we obtain
\begin{equation}
\frac{\dot{a}}{a}=-\frac{2\dot{M_1}}{{M_1}}\left(1-q_1\right),
\label{eq:a_0}
\end{equation}
where $q_1=M_1/M_2$.
When the primary star loses its mass ($\dot{M}_1<0$), thus the orbital separation shrinks (widens) 
for $q_1>1$ ($q_1<1$).
As an example of non-conservative MT, we consider that the primary transfers its mass 
to the secondary ($\dot{M}_1<0$), and a fraction $\beta$ of the transferred mass can accrete on to 
the secondary and a fraction $(1-\beta)$ of the mass is lost from the binary system with a specific orbital
angular momentum $(M_2/M)^2\sqrt{GMa}$.
Then, the orbital evolution is described as
\begin{equation}
\frac{\dot{a}}{a}=-\frac{2\dot{M}_1}{M_1}\left[1-q_1 + \frac{q_1(1-\beta)}{2(1+q_1)}\right].
\label{eq:a_1}
\end{equation}
This type of non-conservative MT would occur when the accretor, 
which is fed by gas accretion, is a compact object 
because the accreted gas onto it releases
a huge amount of energy as radiation and/or outflows and 
a significant fraction of the gas can be ejected \citep{2005ApJ...628..368O,
2014ApJ...796..106J,2015MNRAS.447...49S}.

\subsubsection{mass transfer}
\label{sec:MT}

When one of the stars in a binary exhausts hydrogen at the central core, the star begins to expand.
If the Roche lobe around the primary is filled with its bloated envelope, the material is transferred 
to the secondary star through the first Lagrangian point.
The process is the so-called Roche lobe overflow (RLOF).
The radius of the Roche lobe (Roche radius) is approximately expressed as \citep{Eggleton_1983}:
\begin{equation}
R_{\rm L,1}\simeq a\frac{0.49q_1^{2/3}}{0.6q_1^{2/3}+\ln(1+q_1^{1/3})}.
\end{equation}
The behavior of the MT is determined by the response of the Roche radius and 
the stellar radius when it loses the material \citep{Paczynski_1976}.
Assuming the MT is conservative ($\beta =1$), 
the response of the Roche radius is characterized by
\begin{equation}
\zeta_{\rm L}\equiv \dfrac{d\ln R_{\rm L,1}}{d\ln M_1}\simeq 2.13 q_1-1.67,
\label{eq:zeta_L}
\end{equation}
\citep{Tout_1997}.
For $\zeta_{\rm L}<\zeta_\ast (\equiv d\ln R_1/d\ln M_1)$, the stellar radius shrinks and 
becomes smaller than the Roche radius after the mass of the primary is transferred. 
On the other hand, for $\zeta_{\rm L}>\zeta_\ast$, the MT would proceed unstably and 
the two stars would either merge or experience a common envelope (CE) phase.
The response of the stellar radius is automatically calculated in the stellar evolution calculations, 
while it is estimated with an analytical expression in our semi-analytical model
(see \S\ref{sec:stability}).

We adopt the binary module in \texttt{MESA} to treat the MT process.
When the Roche radius is filled with stellar material ($R_1>R_{\rm L,1}$), 
the mass transfer rate is calculated by a method proposed by \citet{Kolb_1990}, 
\begin{align}
\dot{M}_1&=-2\pi F_1(q_2)\frac{R_{\rm L,1}^3}{GM_1}\\\nonumber
&\times
\left[ \left(\frac{k_{\rm B}T_{\rm ph}}{m_{\rm p}\mu_{\rm ph}}\right)^{3/2}
\frac{\rho_{\rm ph}}{\sqrt{e}}
+\int^{P_{\rm L,1}}_{P_{\rm ph}}F_2(\Gamma_1)
\left(\frac{k_{\rm B}T}{m_{\rm p}\mu}\right)^{1/2}dP,
\right]
\end{align}
where $F_1(q_2)=1.23+0.5\log q_2$, $F_2(\Gamma_1)=\Gamma_1^{1/2}[2/(\Gamma_1+1)]^{(\Gamma_1+1)/(2\Gamma_1-2)}$,
$\Gamma_1$ is the first adiabatic exponent, and $T_{\rm ph}$, $\mu_{\rm ph}$, $\rho_{\rm ph}$, $P_{\rm ph}$ and $P_{\rm L,1}$
are the temperature, mean molecular weight, density and pressure at the photosphere and the radius of $R_{\rm L,1}$, respectively.
The first and second terms correspond to the mass transfer rate for isothermal atmosphere (optically thin) and 
adiabatic atmosphere (optically thick), respectively.
This rate is calculated self-consistently by considering properties of the donor star.
Note that conservative mass transfer, i.e. $\dot{J}_{\rm orb}=0$ and $\dot{M}=0$ ($\beta=1$) is assumed
in our stellar evolution calculations.

\subsubsection{gravitational wave emission}
\label{sec:GW}

A binary loses the orbital angular momentum and the energy due to 
emission of GWs following 
\begin{equation}
\frac{\dot{J}_{\rm orb}}{J_{\rm orb}}=-\frac{32G^3M_1M_2M}{5c^5a^4},
\label{eq:adot_gw}
\end{equation}
where $e=0$ is assumed.
From Eq. (\ref{eq:adot_gw}), the coalescence timescale due to GW emission 
is estimated as 
\begin{align}
t_{\rm GW}&=\frac{5a^4c^5}{256G^3M_1M_2M_{\rm tot}},\nonumber\\
&\simeq 9.5~ \frac{2q_1^2}{1+q_1}\left(\frac{a}{0.2~{\rm AU}}\right)^4
\left(\frac{M_1}{30~\msun}\right)^{-3}~{\rm Gyr}.
\label{eq:tgw}
\end{align}

\subsection{Semi-analytical model}
\label{sec:semiana}

We describe the treatment of PopIII binary stars in our semi-analytical model.
The differences from the detailed calculations of stellar structure are shown in the following.
Since we do not follow the stellar evolution of binaries, instead we need to treat 
single stellar evolution (\S\ref{sec:semiana_star}), the MT rate, 
and the response of the donor and accretor (\S\ref{sec:stability}$-$\ref{sec:MT_term}).
The orbital evolution and GW emission from PopIII BBHs are calculated 
in the same way as in \S\ref{sec:orbit} and \ref{sec:GW}.

\subsubsection{single star evolution}
\label{sec:semiana_star}

We adopt fitting formulae for stellar radii and He core mass of PopIII stars 
with masses of $10~\msun \leq M \leq 100~\msun$
as functions of the mass and time since the birth of the stars \citep{K14}\footnote{The fitting formula 
for the lifetime in the He-burning and He-shell burning phase were updated \citep{2016arXiv161000305K},
which we adopt in this paper.}.
These are based on the results of the stellar evolution calculations for single PopIII stars by \cite{Marigo_2001},
where mass loss due to the stellar wind and pulsation are not considered.
Note that the time-averaged root-mean-square errors of the fitting formulae are within $6\%$ over
the entire lifetime of PopIII stars.
We set the critical mass required for BH formation to $M\geq 28~\msun$.

\subsubsection{stability of mass transfer}
\label{sec:stability}

In our semi-analytical model, we evaluate stability of MT and 
estimate the transfer rate, instead of conducting stellar evolution calculations (\S\ref{MESA}).
We obtain quantitative criteria for the stability of MT.
In summary, PopIII binaries experience unstable MT and CE phases 
if any of the following conditions are satisfied during the evolution;\vspace{2mm}
\\
~~(A) either $M_1$ or $M_2$ exceeds $60~\msun$,\\
~~(B) $M_2/M_1\leq 1/3$ at the first MT episode, and\\
~~(C) $\dot{M}_2>\dot{M}_{\rm crit}~(=2\times 10^{-2}~\msunyr)$.
\vspace{2mm}
\\
In what follows, we explain these conditions in more detail.

(A) {\it Unstable MT due to convective envelope} ---
The stability of MT is determined by the response of the Roche radius $\zeta_{\rm L}$
and the stellar radius of the donor $\zeta_\ast$ (\S\ref{sec:MT}).
In our semi-analytical model, we adopt the analytical expressions for the value of $\zeta_\ast$
as in population synthesis calculations \citep[e.g.,][]{Hurley_2002,2008ApJS..174..223B}.

At the early stage of the MT, mechanical equilibrium is restored 
in a dynamical timescale and the star can be assumed to be at hydrostatic equilibrium at every time.
That is, the response of the stellar radius occurs adiabatically.
The value of $\zeta_{\rm ad}(\equiv [d\ln R_{\rm 1}/d\ln M_1]_S)$ 
depends on whether the stellar envelope is convective or radiative.
For a star with a core and a convective envelope,
the value of $\zeta_{\rm ad,conv}$ is expressed as a function of 
the mass ratio $m_{\rm c,1}$ between the core mass 
and the total mass in the primary
\citep{Hjellming_1987,Soberman_1997}.
The value of $\zeta_{\rm ad,conv}$
increases with $m_{\rm c,1}$ and approaches $-1/3$ for $m_{\rm c,1}\rightarrow 0$, 
which corresponds to a fully convective star.
As shown in Fig.~\ref{fig:HR}, PopIII stars with $M_1>50~\msun$ 
have deep convective envelopes after the onset of He-core burning.
From Table 4 of \cite{Marigo_2001}, 
$m_{\rm c,1}\simeq 0.45-0.5$ 
for $60\la M_1/\msun \la100$, respectively.
Thus, $\zeta_{\rm ad,conv}\simeq 0.44-0.56$ for the mass range,
while $\zeta_{\rm L}=2.13q_1-1.67\geq 0.46$ ($q_1\geq1$, see Eq. \ref{eq:zeta_L}).
Therefore, PopIII binaries which have components heavier than $60~\msun$
are likely to experience unstable MT unless the binary mass ratio is exactly unity or 
the MT occurs before the donor star reaches the Hayashi track because of a small initial separation.
Here, to give a conservative argument, we simply impose that 
{\it if either $M_1$ or $M_2$ exceeds $60~\msun$,
the binary does not form a BBH because of unstable MT}.

(B) {\it Delayed dynamical instability (DDI)} ---
For a star with a radiative envelope, the value of $\zeta_{\rm ad,rad}$ is generally positive.
When the radiative envelope, which has a positive entropy gradient, is removed by MT, 
the layer with a lower entropy is exposed.
Then, the density will increase to adjust its new hydrostatic equilibrium state with the same 
external pressure as before because $(\partial S/\partial \rho)_p=-(1/\rho^2)(\partial T/\partial p)_S<0$.
Thus, the donor's radius shrinks in the dynamical timescale\footnote{
The entropy hardly changes by emitting radiation within a dynamical time
since the thermal relaxation timescale of most parts of the envelope is 
longer than the dynamical timescale.}.
However, the situation in this case is more complicated.
Even if the MT occurs stably at the early phase ($\zeta_{\rm L}<\zeta_{\rm ad,rad}$),
the inner layer of the donor with a shallower entropy gradient can be exposed.
If the entropy profile is shallow enough for the response of the stellar radius to be 
$\zeta_{\rm L}>\zeta_{\rm ad,rad}$,
the MT would become unstable later due to the so-called 
delayed dynamical instability (DDI) \citep{Hjellming_1987}.
The critical mass ratio that leads to the DDI has been estimated as $q_{\rm crit,1}=2-4$
for many kinds of donor stars \citep[e.g.][]{Hjellming_1989,Ge_2010,Pavlovskii_2015,Ge_2015}.
We note that the exact value of the critical mass ratio is still uncertain
(e.g. treatment of super-adiabatic layer, see discussions in \citealt{Pavlovskii_2015}),
in particular, for metal-poor stars like PopIII stars.
We here simply assume $q_{\rm crit,1}=3$, which is 
the averaged value for PopI cases.
Therefore, we assume that {\it if the donor star is sufficiently massive ($M_2/M_1 \leq 1/3$), 
the MT will be unstable due to the DDI}.

(C) {\it Expansion of accretors due to MT} ---
The fitting formulae we adopt are obtained by calculations of 
single stellar evolution, where the stellar structure is thermally relaxed.
However, when the first episode of MT occurs, the secondary (main-sequence) star is away from
its thermal equilibrium state.
In fact, the main-sequence accretor could be bloated, depending on the MT rate.
According to detailed stellar evolution calculations by \cite{Hosokawa_2016}, 
the accreting PopIII stars are unlikely to expand as long as the accretion rate is $\la 10^{-2}~\msunyr$.
The critical rate has been also estimated as a few $\times 10^{-2}~\msunyr$ \citep{Hosokawa_2012}.
The exact critical value is still unclear because it depends on treatment of the boundary 
conditions at the stellar surface.
Here, {\it we adopt the critical accretion rate of $\dot{M}_{\rm crit}\simeq 2\times 10^{-2}~\msunyr$,
above which the accretor would expand, fill its Roche lobe and might lead to unstable MT}.

\begin{figure}
\begin{center}
\includegraphics[width=83mm]
{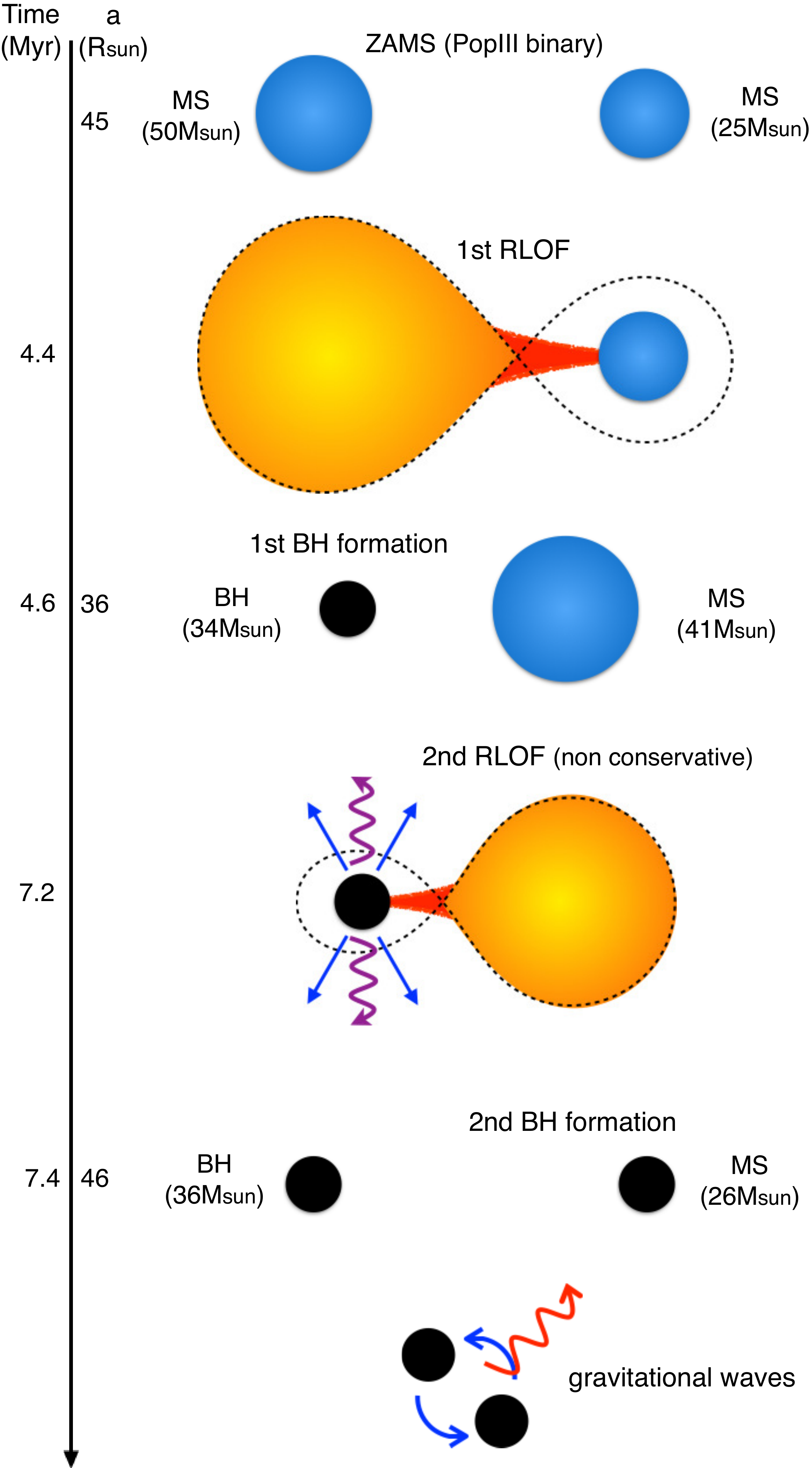}
\caption{Schematic overview of a typical pathway to form 
PopIII BBH without experiencing CE phases.
The initial conditions are set to $M_{1,0}=50~\msun$, $M_{2,0}=25~\msun$ and $a_0=45~\rsun$.
}
\label{fig:schematic}
\end{center}
\end{figure}

\subsubsection{mass transfer rate}

In the following, we focus on cases with stable MT in the dynamical timescale 
($\zeta_{\rm L}<\zeta_{\rm ad}$).
Even in this case, the donor star gradually approaches its thermal equilibrium state,
in which the response of the stellar radius can be satisfied with 
$\zeta_{\rm L}>\zeta_{\rm th}\equiv (d\ln R_1/d\ln M_1)|_{\rm th}$
depending on the stellar structure.
When the Roche lobe is filled with the stellar material due to thermal relaxation, 
MT occurs again in the Kelvin-Helmholtz timescale ($t_{\rm KH}\equiv GM_1^2/R_1L_1$).
The MT rate is given by the overfilling of the donor's Roche lobe, 
$\Delta R_1 \equiv R_1 - R_{\rm L,1}(\geq 0)$, as
\begin{equation}
\dot{M}_1 = -\frac{f(\mu) M_1}{\sqrt{R_1^3/GM_1}}\left(\frac{\Delta R_1}{R_1}\right)^{3}d_{3/2},
\label{eq:MTrate_PS}
\end{equation}
where 
\begin{equation}
f(\mu)=\frac{4\mu\sqrt{\mu}\sqrt{1-\mu}}{(\sqrt{\mu}+\sqrt{1-\mu})^4}\left(\frac{a}{R_1}\right)^3,
\end{equation}
$\mu=M_1/(M_1+M_2)$ and $d_{3/2}=0.2203$
\citep[e.g.][]{Paczynski_1972,Savonije_1978,Edwards_Pringle_1987}.
Note that a polytropic equation of state with an index of $n=3/2$ is assumed 
in Eq. (\ref{eq:MTrate_PS}).
In fact, radial dependence of the adiabatic index affects the MT rate 
\citep[see e.g., ][]{Kolb_1990,Ge_2010}.
However, we here adopt the simpler prescription because our semi-analytical model has a lack of 
thermal properties of the donor's envelope.

The MT rate in Eq.(\ref{eq:MTrate_PS}) is high soon after the onset of the MT.
However, the MT proceeds in the KH timescale of the donor star.
Thus, we set the maximum MT rate to $\dot{M}_{\rm 1,max}=M_1/t_{\rm KH,1}$.
Typically, $t_{\rm KH,1}\simeq 4\times 10^3$ yr $(M_1/50~\msun)^2(R_1/20~\rsun)^{-1}(L_1/10^6~\lsun)^{-1}$
and then $\dot{M}_{\rm 1,max}\simeq 1.25\times 10^{-2}~\msunyr$.
To estimate the KH timescale, we adopt a fitting form of the stellar luminosity at the beginning 
of He-shell burning phase given by \cite{K14}.
In fact, since the stellar luminosity of the donor rapidly decreases during the MT \citep{Paczynski_1967},
we underestimate the value of $t_{\rm KH,1}$ and overestimate the maximum accretion rate.
In other words, the condition (C) in \S\ref{sec:stability} would be alleviated somewhat in 
detailed stellar evolution calculations.

\begin{figure}
\begin{center}
\includegraphics[width=78mm]
{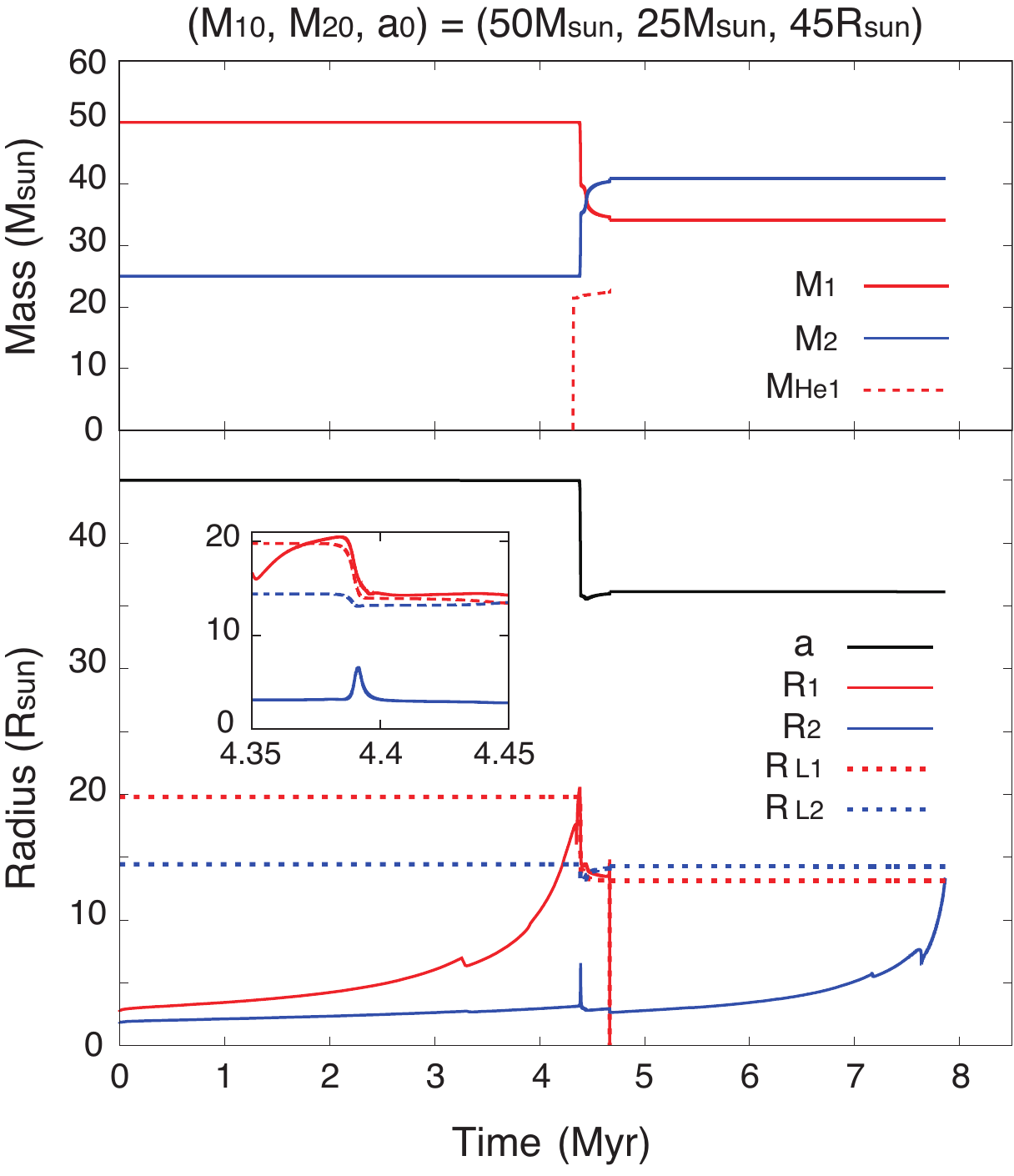}
\caption{
{\it Top}: Time evolution of stellar masses of a PopIII binary
obtained by stellar evolution calculation with {\tt MESA}.
The initial conditions are set to $M_{1,0}=50~\msun$, $M_{2,0}=25~\msun$ and $a_0=45~\rsun$.
The primary (red) and secondary (blue) mass and the He core mass of the primary star (red dashed) are shown.
{\it Bottom}: 
Time evolution of stellar radii (red and blue solid) and 
the orbital separation (black solid) of the same binary as in the top panel.
The Roche radii of the two stars are shown by dashed curves.
After the MT, the binary properties change to $M_1=34~\msun$, $M_2=41~\msun$, 
$M_{\rm He,1}=22~\msun$ and $a=36~\rsun$.
We set the He core mass to zero after the star collapses into a BH.
}
\label{fig:50-25M_45R}
\end{center}
\end{figure}

\begin{figure}
\begin{center}
\includegraphics[width=78mm]
{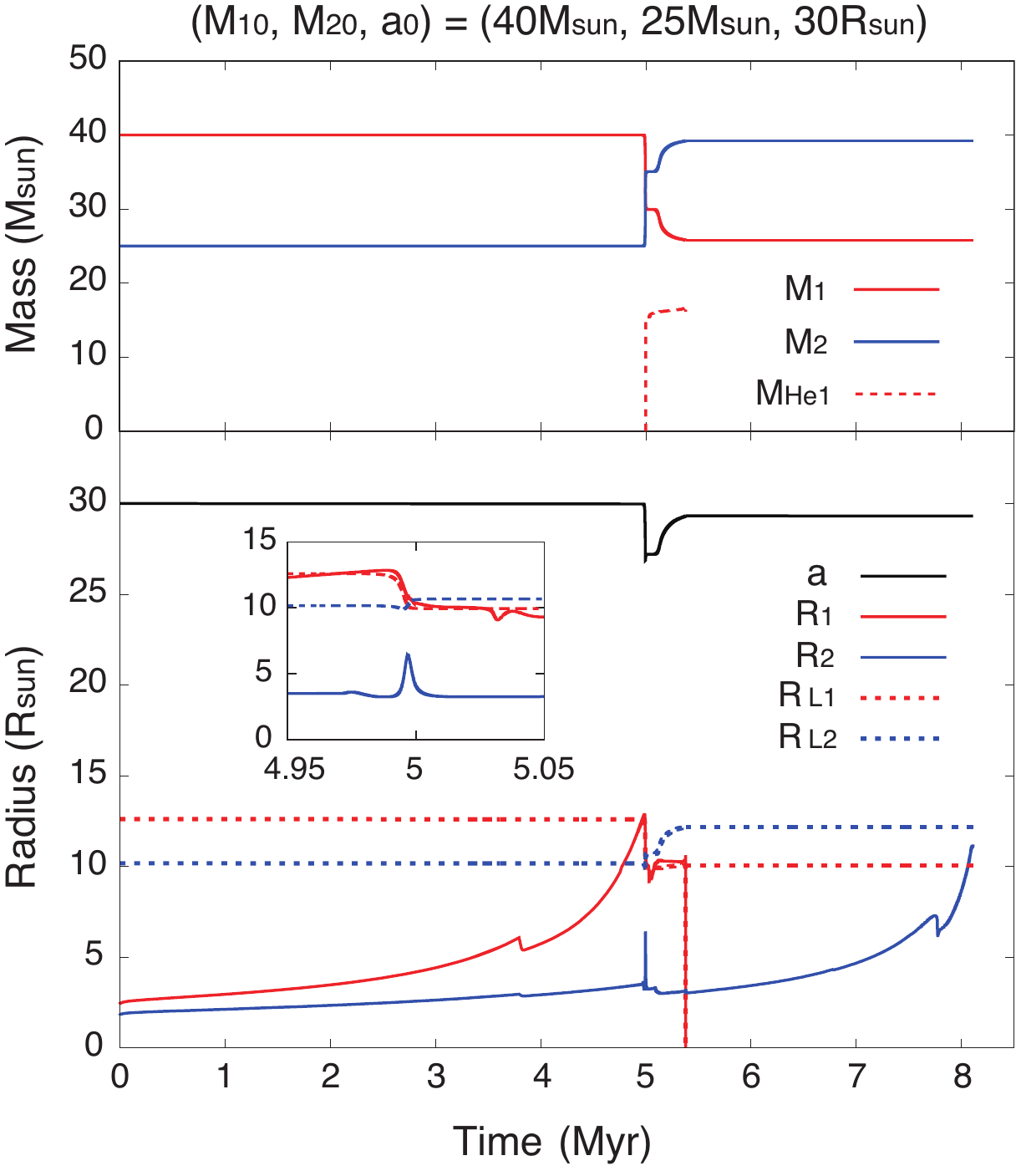}
\caption{
Same as Fig.~\ref{fig:50-25M_45R},
but different initial conditions of the binary;
$M_{1,0}=40~\msun$, $M_{2,0}=25~\msun$ and $a_0=30~\rsun$.
After the MT, the binary properties change to $M_1=26~\msun$, $M_2=40~\msun$, 
$M_{\rm He,1}=16~\msun$ and $a=29~\rsun$.
}
\label{fig:40-25M_30R}
\end{center}
\end{figure}

\subsubsection{stellar radii during MT and rejuvenation}
\label{sec:rej}

During a MT episode, the donor's radius changes depending on 
how much of the mass is removed from the envelope.
This process is generally complicated and difficult to be treated properly.
For simplicity, we consider cases where the donor star is a giant 
in which He burning begins and the accretor is a main-sequence star.
Note that MT between two main-sequence stars (the so-called case A) is not considered
to avoid complicated situations.

The stellar radius of the donor during the MT is estimated 
using the fitting formulae for single star models
as $R_1(M_1,t,t_i(M_1))\rightarrow R_1(M_1+\delta M_1,t+\delta t,t_i(M_1))$,
where $\delta t$ is the time after the onset of the MT, 
$\delta M_1=\int _0^{\delta t}\dot{M}_1dt~(<0)$ and 
$t_i(M_1)$ is the lifetime during 
He core ($i={\rm He}$) and He-shell ($i={\rm HeS}$) burning phases.
This prescription assumes that the core mass and the stellar age 
in the corresponding phase does not change during the MT
because the MT occurs faster than the nuclear core and/or shell burning 
(see also the results with {\tt MESA} shown in Figs.~\ref{fig:50-25M_45R} and \ref{fig:40-25M_30R}).

The accretor is fed at a rate of $\dot{M}_2(=-\dot{M}_1>0)$.
Since unburnt hydrogen is supplied as fuel,
we modify the stellar age (i.e. rejuvenation) following a method suggested by
\cite{Tout_1997} and \cite{Hurley_2002}.
We approximate that the mass of H-burning core is proportional to $M_2\cdot t/t_{\rm H}(M_2)$,
where $t$ is the stellar age.
Thus, we estimate a new stellar age due to rejuvenation as
\begin{equation}
t'=\frac{M_2}{M_2+\delta M_2}\frac{t_{\rm H}(M_2+\delta M_2)}{t_{\rm H}(M_2)}t.
\label{eq:new_t}
\end{equation}
Using Eq. (\ref{eq:new_t}), we estimate the stellar radius of the accretor as
$R_2(M_2+\delta M_2, t', t_{\rm H}(M_2+\delta M_2))$.

\subsubsection{termination of MT}
\label{sec:MT_term}

The episode of MT terminates at the end of lifetime of the donor or
when a large fraction of its hydrogen envelope is stripped from the donor.
In the latter case, the donor's radius rapidly shrinks when 
the He core mass dominates its total mass.
Note that we define the core boundary as the outermost part 
where the H mass fraction is lower than $0.01$ and 
the He mass fraction is higher than $0.1$.
The critical ratio of the He core to the total mass is supposed to be $q_{\rm He,crit}\simeq 0.6-0.8$, 
above which the donor star evolves blueward in the HR diagram \citep{Kippenhahn_1990}.
In our semi-analytical model, we set $q_{\rm He,crit}\simeq 0.58$
so that the evolutionary tracks of the PopIII binaries agree with 
the results by the stellar evolution calculations.


\section{Population III binary BH formation}
\label{sec:PopIIIBBH_result}

In this section, we discuss the evolutionary paths of PopIII binaries. As an example, 
Fig.~\ref{fig:schematic} shows a schematic picture of the time evolution of a PopIII binary
with the initial masses of $M_{1,0}=50~\msun$ and $M_{2,0}=25~\msun$ and the initial separation of $a_0=45~\rsun$.
This is a typical pathway of formation of PopIII BBHs without unstable MT and CE phases.
In what follows, we first demonstrate two cases of binary evolution pathways calculated with {\tt MESA}.
Then, we compare these results with the semi-analytic formulae developed in the previous section.

\begin{figure}
\begin{center}
\includegraphics[width=77mm]
{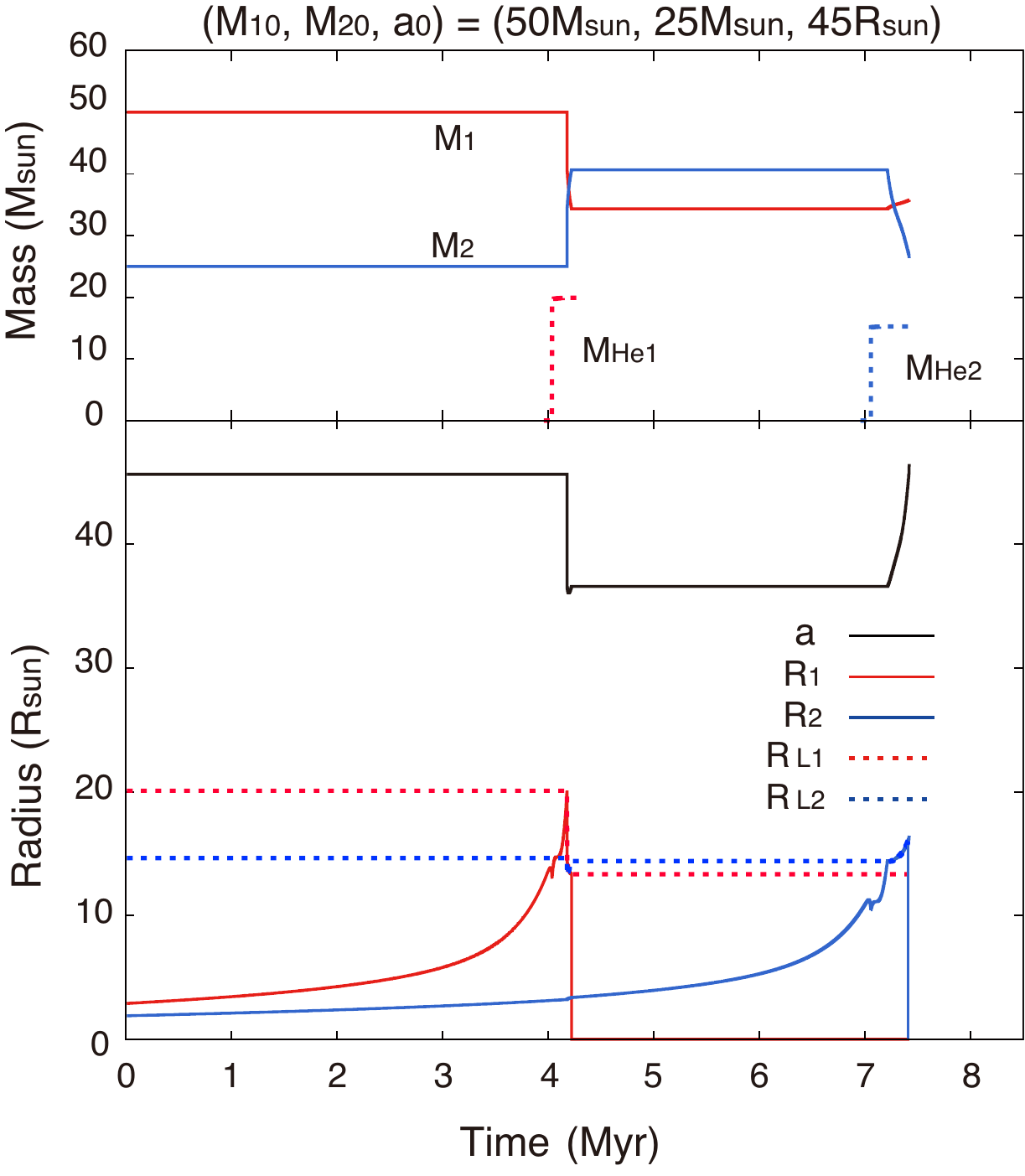}
\caption{
Time evolution of stellar masses, stellar radii, and the orbital separation of a PopIII binary calculated by
the semi-analytical model.
The initial conditions are set to $M_{1,0}=50~\msun$, $M_{2,0}=25~\msun$ and $a_0=45~\rsun$,
which are the same as in Fig.~\ref{fig:50-25M_45R}.
}
\label{fig:semi_50_25-45}
\end{center}
\end{figure}

\begin{figure}
\begin{center}
\includegraphics[width=77mm]
{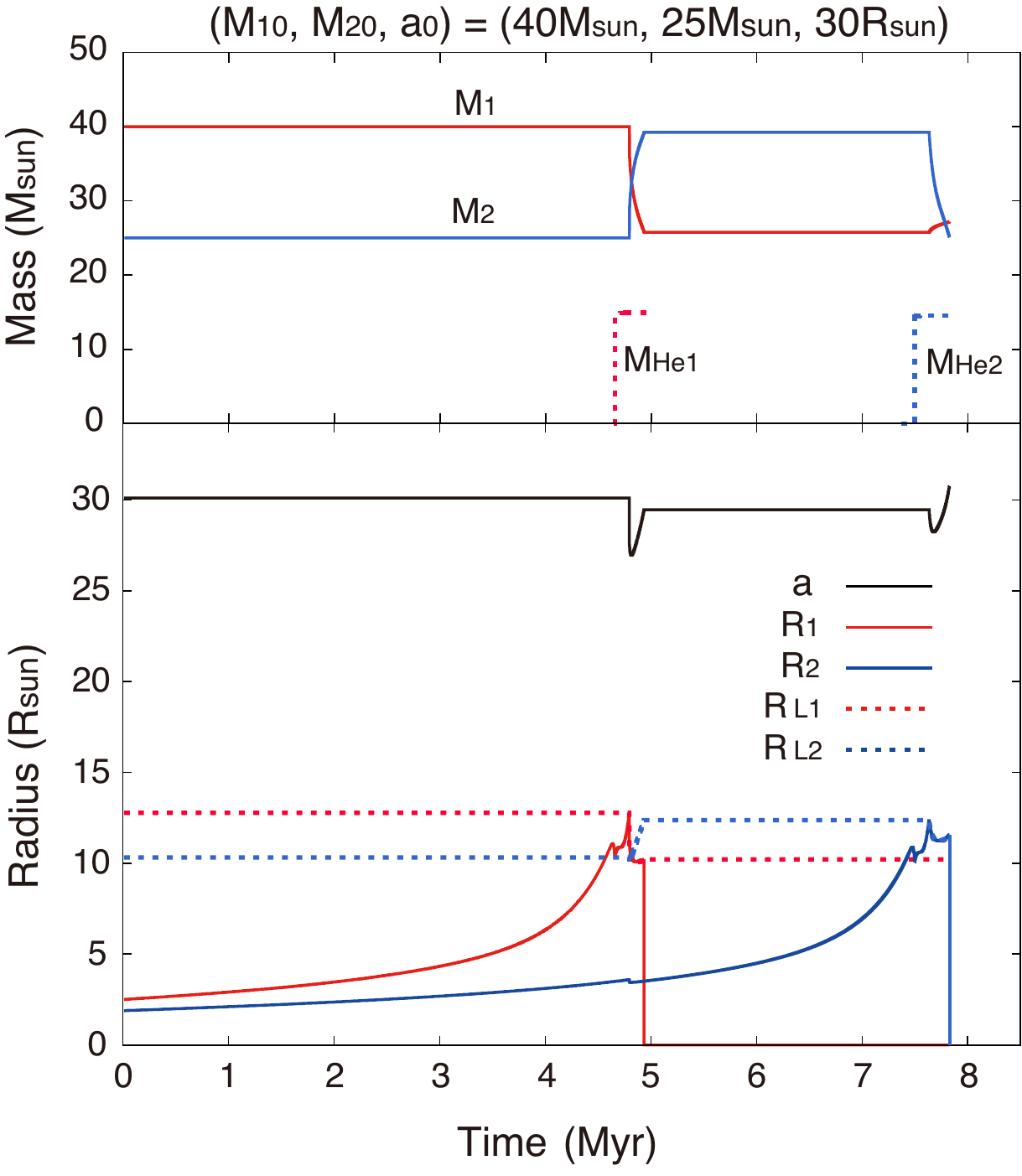}
\caption{Same as Fig.~\ref{fig:semi_50_25-45} (semi-analytical model),
but different initial conditions of the binary;
$M_{1,0}=40~\msun$, $M_{2,0}=25~\msun$ and $a_0=30~\rsun$,
which are the same as in Fig.~\ref{fig:40-25M_30R}.
}
\label{fig:semi_40_25-30}
\end{center}
\end{figure}

\subsection{Evolution of individual PopIII binaries}

\subsubsection{stellar evolution calculations}
\label{sec:result_sec}

We present the time evolution of two PopIII binaries with 
different initial conditions of $(M_{1,0}, M_{2,0}, a_0) = (50~\msun, 25~\msun ,45~\rsun)$
in Fig.~\ref{fig:50-25M_45R} and 
$(M_{1,0}, M_{2,0}, a_0) = (40~\msun, 25~\msun ,30~\rsun)$
in Fig.~\ref{fig:40-25M_30R}, respectively.
Red and blue curves show the evolution of the primary and secondary star, respectively.
In each figure, the top panel shows the mass evolution of the two stars $M_{1(2)}$ (solid)
and the He core of the primary star $M_{\rm He,1}$ (dashed),
and the bottom panel shows the evolution of the stellar radii $R_{1(2)}$ (solid), 
the Roche radii $R_{\rm L,1(2)}$ (dashed) and the orbital separation $a$ (black solid).
Note that these results shown in Figs.~\ref{fig:50-25M_45R} and \ref{fig:40-25M_30R}
are obtained by stellar evolution calculations with {\tt MESA}.
We follow the evolution of these binaries until just before the second episode of MT occurs
because the MT (from the secondary star to a BH) is not conservative.
Instead, the non-conservative MT is studied with our semi-analytical model,
exploring the dependence of the results on the parameter $\beta$ (see Appendix A).

As shown in Fig.~\ref{fig:50-25M_45R},
after the two stars evolve from their ZAMS to post-main-sequence phases,
their stellar radii expand.
The primary star fills its Roche lobe after the end of the H-core burning
and the first episode of MT begins at $t\simeq 4.39$ Myr.
Since the mass ratio at the onset of the MT is larger than unity ($q_1>1$) 
and the MT is  conservative ($\dot{M}_1=-\dot{M}_2>0$), 
the separation initially shrinks (see Eq. \ref{eq:a_0}).
After the mass ratio becomes smaller than unity, the MT continues and thus the separation gets wider.
Once the hydrogen rich layer of the envelope is removed, the MT terminates at $t\simeq4.6$ Myr 
due to the discrete change in the surface composition. 
This occurs just beneath the outermost convection zone in the hydrogen shell, 
where the helium abundance increases by a factor of $\sim 2$. 
By this time, the He core mass occupies $\sim 65\%$ of the total mass.
The core mass of $\simeq 22~\msun$ is massive enough to form a BH by direct collapse.
Note that we do not calculate the evolution of the (primary) naked He star
but the primary star is considered as a BH with $M_1\simeq 22~\msun$
because of the absence of the wind-mass loss.

During the first episode of MT, the mass of the secondary star is increased to $\simeq 41~\msun$.
The maximum MT rate  is $\simeq 2\times 10^{-3}~\msunyr$ at the early stage
and drops to $\la 10^{-5}~\msunyr$ later.
After the MT, the secondary star becomes massive enough to form a BH.

The overall behaviors for the two cases shown in 
Figs.~\ref{fig:40-25M_30R} and \ref{fig:50-25M_45R} are similar.
For both cases, the first episodes of MT terminate when the He-core mass of the primary 
stars occupies $65~(62)\%$ of their total mass.

\subsubsection{semi-analytical model}
\label{sec:semiana_result}

In Figs.~\ref{fig:semi_50_25-45} and \ref{fig:semi_40_25-30},
we show the time evolution of two PopIII binaries calculated by the semi-analytical model.
The same initial conditions are adopted for both cases as in Figs.~\ref{fig:50-25M_45R} 
and \ref{fig:40-25M_30R}.
The evolutionary tracks until just before the second episodes of MT 
are very similar to those of the stellar evolution calculations.
Note that our semi-analytical model does not replicate expansion phases of the 
secondary stars during the first episodes of MT as shown in Figs.~\ref{fig:50-25M_45R} 
and \ref{fig:40-25M_30R}.
Here, the critical core mass ratio is set to be $q_{\rm He, crit}=0.58$ in order to determine 
the termination of the MT.
This value is slightly smaller than the actual values estimated in \S\ref{sec:result_sec},
but allows us to reproduce the plausible results by our simple model.

In the second episode of MT, the donor is an ordinary star but the accretor is a BH.
Since the MT rate is estimated as $|\dot{M}_2|\sim 10^{-5}-10^{-3}~\msunyr$, 
the BH is fed at a super-Eddington accretion rate of $|\dot{M}_2|/\dot{M}_{\rm Edd,1}\simeq O(10-10^3)$, 
where $\dot{M}_{\rm Edd,1}\equiv L_{\rm Edd,1}/(\epsilon c^2)\simeq 6.9\times 10^{-7}
~\msunyr (M_1/30~\msun)(\epsilon/0.1)^{-1}$, $L_{\rm Edd,1}$ is the Eddington luminosity
and $\epsilon$ is the radiative efficiency.
Although one assumes that the accretion rate is limited by the Eddington accretion, 
i.e. $\dot{M}_1={\rm min}(|\dot{M}_2|,~\dot{M}_{\rm Edd,1}$), super(hyper)-Eddington accretion would be possible
through an optically-thick accretion disk associated with outflows and/or jets
\citep[e.g.][]{1978MNRAS.184...53B,1988ApJ...332..646A,2005ApJ...628..368O,
2014ApJ...796..106J,2015MNRAS.447...49S}.
We simply assume that a fraction $\beta=\dot{M}_1/|\dot{M}_2|=0.1$ 
of the gas can accrete onto the BH 
and the rest is ejected from the BH with a certain specific angular momentum.
We follow the orbital evolution using Eq. (\ref{eq:a_1}).
Finally, the stellar envelope of the secondary star is stripped and the second episode of MT terminates.
The secondary (naked He star) collapses into a BH and thus a binary BH system forms.

\subsection{Parameter dependence on final states of PopIII binaries}

\begin{figure}
\begin{center}
\includegraphics[width=82mm]{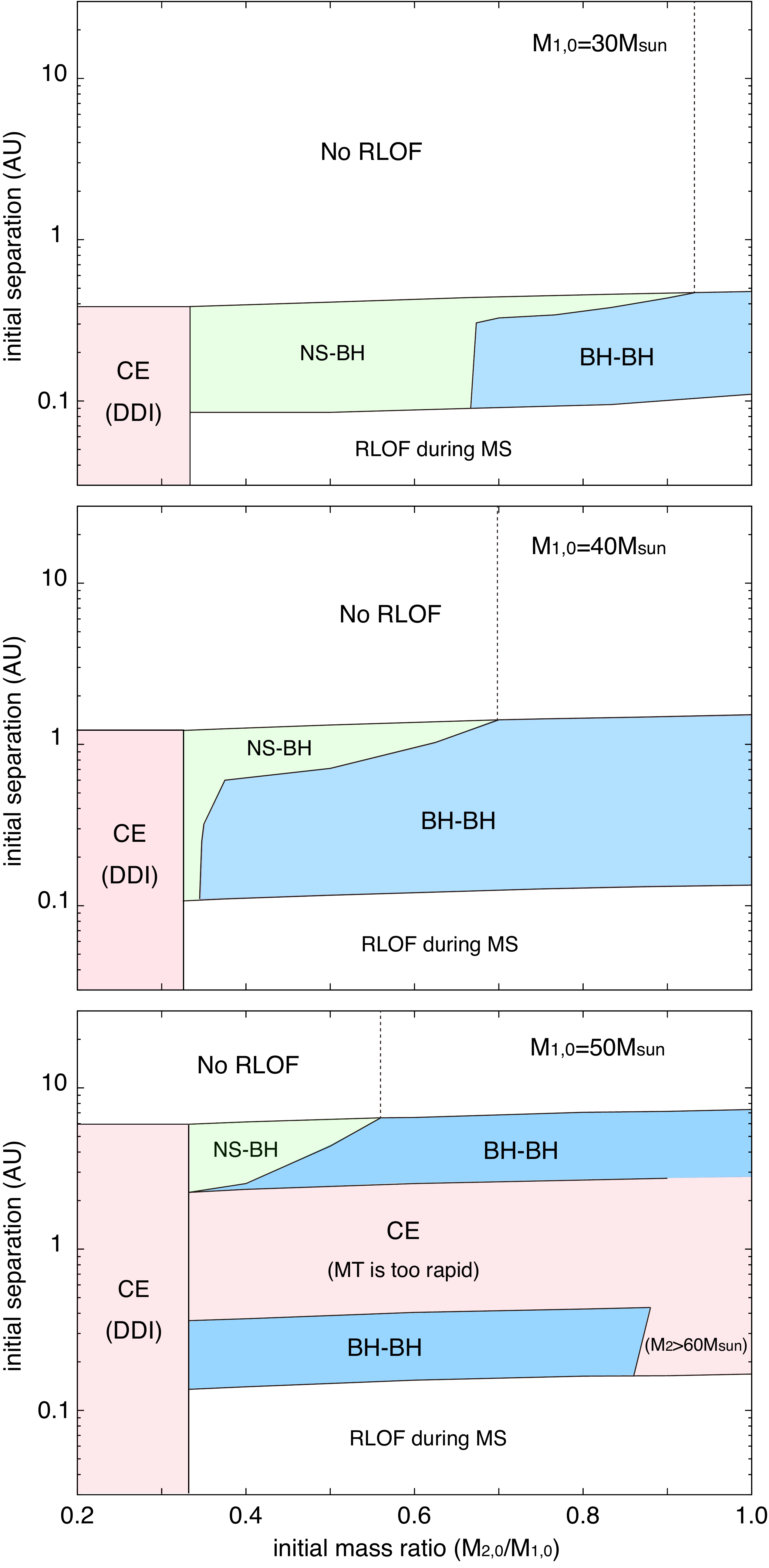}
\end{center}
\caption{Dependence of final states of PopIII binaries on the initial conditions of the mass ratio 
$q_{2,0}(=M_{2,0}/M_{1,0})$ and the orbital separation $a_0$.
Shaded regions indicate conditions for which BBHs form (blue), 
NS-BH form (green) and PopIII binaries experience CE phases (red).
For $q_{2,0}\leq 1/3$, the binaries experience the delayed dynamical instability (DDI).
For $M_{1,0}=50~\msun$, the binaries are likely to experience CE phases 
during either the first or second episode of MT.
}
\label{fig:M1_q_a}
\end{figure}

Fig.~\ref{fig:M1_q_a} shows final fates of PopIII massive binaries with different 
initial conditions of $q_{2,0}=M_{2,0}/M_{1,0}$ and $a_0$
for three different primary masses of $M_{1,0}=30$ (top), $40$ (middle) and $50~\msun$ (bottom), respectively.
Shaded regions indicate initial conditions for which massive PopIII binaries form BBHs (blue),
form NS-BH binaries (green) and experience CE phases due to unstable MT (red).
With larger initial separations, stellar radii of the two stars never exceed their Roche radii, 
and they do not interact through MT, indicated as ``No RLOF", 
where each of the stars evolve as a single star.
Dashed lines show boundaries, in the right-hand side of which the binary can form a wide-separation BBH 
even without binary interactions.
It is worthy noting that such wide-separation BBHs formed in ``No RLOF" do not merge within the Hubble time. 
On the other hand, with very small separations, the primary star fills its Roche radius even 
during its main-sequence phase (the so-called case A mass transfer).
In this case, episodes of MT would be likely to affect the evolution of the core mass
because the core evolution has not been decoupled yet from that of the total mass.
Thus, our treatment for the stellar evolution and its rejuvenation would not work.
In addition, the effects due to tidal force could be important for such close binaries 
(see \S\ref{sec:spin}).

In the top panel, the parameter dependence of the final fates of PopIII binaries 
with $M_{1,0}=30~\msun$ is shown.
With smaller mass ratio of $q_{2,0}=M_{2,0}/M_{1.0}\leq 1/3$, the MT is unstable 
by the DDI (see \S\ref{sec:stability}) and thus the binary plunges into a CE phase.
For other cases, compact binaries (BBH or NS-BH) are formed as remnants only via stable MTs.
Even for an initially less massive star, its mass can increase via MT to 
$\ga M_{\rm crit,BH}(\sim 28~\msun)$, above which the star collapses directly into a BH without explosions.
In other words, the MT from the massive primary allows the secondary to be massive enough to form a BH.
For higher initial primary mass ($M_{1,0}=40~\msun$), the overall result is similar to that for $M_{1,0}=30~\msun$,
but BBHs can form for a wider range of initial conditions of binaries.

For the highest primary mass ($M_{1,0}=50~\msun$), 
the PopIII binary is more likely to experience CE phases for 
the two reasons described in \S\ref{sec:stability}.
One is that the secondary star becomes more massive than $\simeq 60~\msun$ via the first episode of MT.
Thus, when the secondary star fills its Roche lobe and undergoes the second episode of MT, 
the MT is unstable because its stellar envelope is convective.
Another reason is that the MT rate from the primary to the secondary can be higher than 
the critical value of $\dot{M}_{\rm crit,2}$,
above which the secondary stellar radius would be bloated.
As a result, in this case ($M_{1,0}=50~\msun$), PopIII binaries with smaller ($\sim 0.2$ AU) 
or lager ($\sim 4$ AU) separations can form BBHs via stable MT without experiencing CE phases.

In summary, we find that there are regions in the parameter space where
merging BBHs are formed from PopIII binaries without experiencing CE phases.
We consider these pathways are robust to form PopIII merging BBHs.
In the following discussion, we focus only on such a population
because the outcome of the CE phases of PopIII binaries is uncertain. 
In this sense, our argument on the event rate 
of BBH mergers is somewhat conservative.


\section{Formation efficiency and population of PopIII binary BHs}
\label{sec:BBHeff}

In the previous section, we discussed formation pathways of PopIII BBHs,
in particular, focusing on those with or without unstable MT and CE phases.
Applying the results, we estimate efficiency of PopIII BBH formation 
for given initial conditions of PopIII binaries.
In the following, we discuss the BBH formation efficiency with an analytical argument in \S\ref{sec:analytic}.
As we will see, this estimate gives reasonable BBH formation efficiencies and
the delay time distribution.
We also build population synthesis models
for two types of binary initial conditions
as examples (see \S\ref{sec:bic} and \ref{sec:bdis}).
The initial conditions of $N_{\rm tot}$ binaries are generated with the Monte Carlo method
and then the evolution is followed by the semi-analytic formula developed in \S\ref{sec:semiana}. 
The convergence of the results has been checked for $N_{\rm tot}=10^5$ and $10^6$.
Note that we focus only on PopIII BBH formation through  stable MT,
i.e., without experiencing any case A MT and CE phases.
If BBHs formed via these processes, the efficiency of BBH formation could be higher.

\subsection{Analytical estimate}
\label{sec:analytic}

Here we estimate analytically the number fraction of PopIII BBHs coalescing within the Hubble time 
to the total number of PopIII binaries, $f_{\rm BBH}=I/J$, where
\begin{equation}
I=\int^{M_{\rm crit,2}}_{M_{\rm crit,1}}dM_1~\Psi (M_1)
\int^{1}_{q_{\rm crit}}dq ~\Phi(q)
\int^{a_{\rm GW}}_{a_{\rm crit}}da ~\Gamma(a),
\label{eq:f_BBH_1}
\end{equation}
and 
\begin{equation}
J=\int^{M_{\rm max}}_{M_{\rm min}}dM_1~\Psi (M_1)
\int^{1}_{q_{\rm min}}dq ~\Phi(q)
\int^{a_{\rm max}}_{a_{\rm min}}da ~\Gamma(a).
\label{eq:f_BBH_2}
\end{equation}
Here $\Psi (M_1)$ is the IMF for the primary star with $M_{\rm min}\leq M_1 \leq M_{\rm max}$,
$\Phi(q)$ is the initial mass-ratio distribution with $q_{\rm min}(=M_{\rm min}/M_1) \leq q \leq 1$, 
$\Gamma(a)$ is the initial distribution of binary separations with $a_{\rm min}\leq a \leq a_{\rm max}$.
For a primary star with $M_1>M_{\rm crit,1}$, it forms a BH by direct collapse.
For $M_1>M_{\rm crit,2}$, the primary evolves a red giant with a deep convective envelope,
resulting in unstable MT and a CE phase (see \S\ref{sec:stability}).
We here set $M_{\rm crit,1}=28~\msun$ and $M_{\rm crit,2}=60~\msun$, respectively.
We define $q_{\rm crit}={\rm max}(q_{\rm min},q_{\rm CE})$, where $q_{\rm CE}=1/3$ is the critical mass ratio 
below which a binary could evolve through a CE phase
due to the delayed dynamical instability.
As shown in \S\ref{sec:semiana_result} (Figs.~\ref{fig:semi_50_25-45} and \ref{fig:semi_40_25-30}), 
the orbital separation after forming a BBH hardly changes from the initial separation.
The critical separation for a formed BBH required to merge within the Hubble time is 
estimated as $a_{\rm GW}\simeq 0.22~{\rm AU}~(M_1/30~\msun)^{3/4}$, 
where formed BBHs are assumed to be equal-mass binaries (see Figs \ref{fig:semi_50_25-45} and \ref{fig:semi_40_25-30}).
The value of $a_{\rm crit}(\simeq 0.1$ AU) is adopted so that PopIII binaries 
do not experience (case A) MT during their main-sequence phases (see Fig.~\ref{fig:M1_q_a}).
As long as $\Gamma(a)$ is not a steep function at $0.1 \la a/{\rm AU} \la 0.3$,
therefore, Eq. (\ref{eq:f_BBH_1}) can be approximated as 
\begin{equation}
I\simeq \int^{a_{\rm GW,0}}_{a_{\rm crit,0}}da ~\Gamma(a) 
\times \int^{M_{\rm crit,2}}_{M_{\rm crit,1}}dM_1~\Psi (M_1)
\int^{1}_{q_{\rm crit}}dq ~\Phi(q),
\label{eq:f_BBH_3}
\end{equation}
where $a_{\rm GW,0}= 0.22$ AU and $a_{\rm crit,0}= 0.1$ AU.
Using Eqs. (\ref{eq:f_BBH_2}) and (\ref{eq:f_BBH_3}), and 
assuming $\Gamma(a)\propto a^\gamma$, we can estimate
\begin{align}
f_{\rm BBH}&\simeq 
\dfrac{\int^{M_{\rm crit,2}}_{M_{\rm crit,1}}dM_1~\Psi (M_1)
\int^{1}_{q_{\rm crit}}dq ~\Phi(q)}
{\int^{M_{\rm max}}_{M_{\rm min}}dM_1~\Psi (M_1)
\int^{1}_{q_{\rm min}}dq ~\Phi(q)}
\nonumber \\
&\times 
\begin{cases}
\frac{\ln (a_{\rm GW,0}/a_{\rm crit,0})}{\ln(a_{\rm max}/a_{\rm min})}~~~~~~(\gamma=-1),\vspace{2mm}\\
\frac{(a_{\rm GW,0}^{\gamma+1} - a_{\rm crit,0}^{\gamma+1})}
{(a_{\rm max}^{\gamma+1} - a_{\rm min}^{\gamma+1})} ~~~~~~~(\gamma \neq -1).
\end{cases}
\label{eq:f_BBH_4}
\end{align}
In addition, we can estimate the number of PopIII BBHs which merge in coalescence timescales of $t_{\rm GW}$ as
\begin{equation}
\frac{dN}{dt_{\rm GW}}= \frac{dN}{da}\frac{da}{dt_{\rm GW}}\simeq 
\Gamma(a)\frac{a}{t_{\rm GW}} 
\propto t_{\rm GW}^{(\gamma-3)/4},
\end{equation}
where we use $dN/da\simeq \Gamma(a)\propto a^\gamma$ and $t_{\rm GW}\propto a^4$ (see Eq. \ref{eq:tgw}).
Note that the coalescence time distribution is not sensitive to the choice of $\gamma$.

It is worthy estimating the efficiency of BBH formation which allows pathways even 
through unstable MT and CE phases.
The efficiency can be roughly estimated as $f_{\rm BBH,max}=K/J$, 
where 
\begin{equation}
K\simeq \int^{a_{\ast,0}}_{a_{\rm crit,0}}da ~\Gamma(a) 
\times \int^{M_{\rm max}}_{M_{\rm crit,1}}dM_1~\Psi (M_1)
\int^{1}_{q_{\rm min}}dq ~\Phi(q).
\label{eq:f_BBH_6}
\end{equation}
Here, the conditions (A) and (B) shown in \S\ref{sec:stability} are removed,
i.e., $M_{\rm crit,2}\rightarrow M_{\rm max}$ and $q_{\rm crit}\rightarrow q_{\rm min}$.
We also allow rapid MT to form BBHs (see condition C in \S\ref{sec:stability}) and thus
set $a_{\ast,0}\simeq 1$ AU (see bottom panel in Fig. \ref{fig:M1_q_a}).

\begin{figure}
\begin{center}
\includegraphics[width=84mm]
{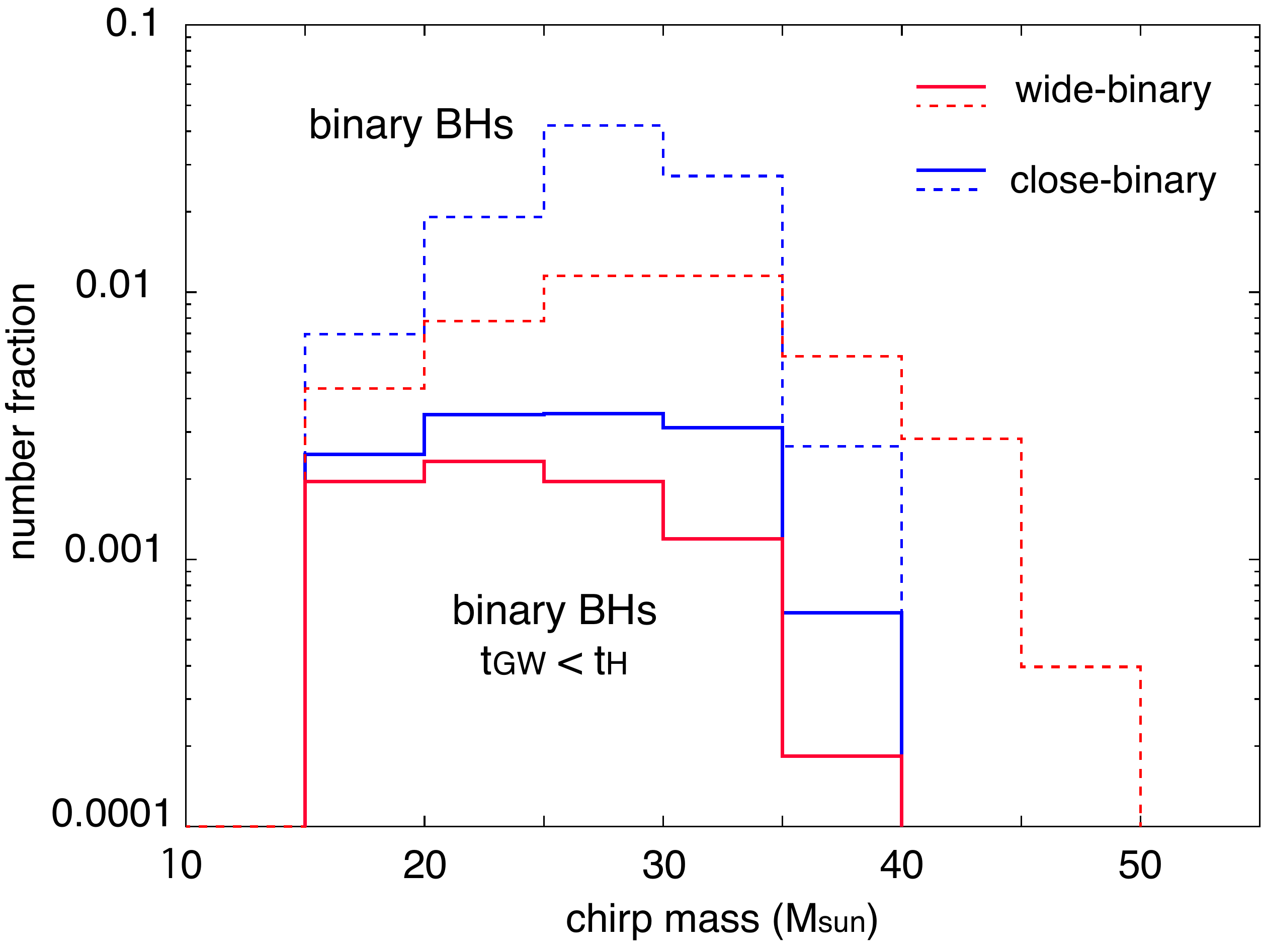}
\caption{Chirp mass distribution of PopIII BBHs for the wide-binary (red) 
and close-binary (blue) model, respectively.
PopIII BBHs formed through stable MT (dashed) and PopIII BBHs coalescing 
within the cosmic age, $t_{\rm GW}< 13.8$ Gyr (solid) are shown.
}
\label{fig:Mch_hist}
\end{center}
\end{figure}

\subsection{Two types of binary initial conditions}
\label{sec:bic}

PopIII binaries would form due to fragmentation in a circumstellar disk at $\sim 1-100$ AU
\citep{2008ApJ...677..813M,2011Sci...331.1040C,2012MNRAS.424..399G,Hosokawa_2016}
and/or at larger scales of $\sim 10^3$ AU \citep{2009Sci...325..601T,2013MNRAS.433.1094S}.
In addition, massive PopIII stars emit strong UV radiation \citep{2011Sci...334.1250H,2012MNRAS.422..290S}, 
which affects gas accretion onto the proto-binary and the final masses.
\cite{2014ApJ...792...32S} have studied statistical properties of PopIII binaries in $\sim 60$ mini-halos,
performing radiation hydrodynamical simulations with a spacial resolution of $\sim 10$ AU.
As a result, the binary fraction is $\sim 50\%$ and the separation is $10-10^3$ AU.
\cite{Stacy_2016} have performed a cosmological simulation 
including radiative feedback with a very high-resolution of $\sim 1$ AU.
They studied star formation in one mini-halo and found that a close massive PopIII binary forms
with $13+ 15~\msun$ and $a\sim 5$ AU.

\subsubsection{wide-binary model}
As an example, we consider the same initial conditions as in \citet{K14},
where a classical model for initial conditions of field binaries \citep[e.g.,][]{Hurley_2002}
is applied for PopIII binaries.
Namely, we adopt a flat IMF, $\Psi (M_1)\propto {\rm const.}$, for the primary star with $10\leq M_1/\msun \leq 100$,
a flat mass-ratio distribution, $\Phi (q)\propto {\rm const.}$, with $q_{\rm min}(=10~\msun/M_1)\leq q\leq 1$,
and a log-flat distribution for the orbital separation, $\Gamma(a)\propto a^{-1}$.
The minimum separation is set so that the binary does not fill its Roche radius from the beginning.
We set $a_{\rm max}=10^6~\rsun$ ($\simeq 4.6\times 10^3$ AU).
In this model, the number fraction of PopIII BBHs formed by stable MT is estimated 
from Eq. (\ref{eq:f_BBH_4}) as 
\begin{equation}
f_{\rm BBH}\simeq \frac{\ln \left(\dfrac{a_{\rm GW,0}}{a_{\rm crit,0}}\right)}
{\ln \left(\dfrac{a_{\rm max}}{a_{\rm min}}\right)}
\times \dfrac{\frac{2}{3}(M_{\rm crit,2}-M_{\rm crit,1})}
{M_{\rm max}-M_{\rm min}[1+\ln(\frac{M_{\rm max}}{M_{\rm min}})]}.
\label{eq:f_BBH_5}
\end{equation}
Note that this equation is valid for $M_{\rm min}\leq M_{\rm crit,1}/3\simeq 9.3~\msun$.
Adopting the values of the initial conditions, we estimate as $f_{\rm BBH}\simeq 0.02$
from Eq. (\ref{eq:f_BBH_5}),
where we adopt $a_{\rm min}\simeq 3R_1\simeq 7~\rsun$.
In this condition, adding to $f_{\rm BBH}\simeq 0.02$ the possibility that BBHs are also formed 
by unstable MT, Eq. (\ref{eq:f_BBH_6}) gives a maximum formation efficiency 
$f_{\rm BBH, max}\sim 0.1$, which is consistent with that by \cite{K14}.

\subsubsection{close-binary model}

Different initial conditions have been suggested by \cite{2016arXiv161201524B} 
for close Pop III binaries (close-binary model). 
These initial conditions are motivated by
N-body simulations by \citet{Ryu_2016},
who follow the time evolution of multiple systems in a star-forming cloud in a mini-halo.
They considered that PopIII stars form due to disk fragmentation around $10-20$ AU
and these stars lose their angular momentum by dynamical friction with the ambient gas,
resulting in formation of stable close binaries\footnote{\citet{Ryu_2016} assumed a uniform density 
distribution of the ambient gas ($n_{\rm gas}=10^6~\cc$) at $10-20$ AU, and neglected radiative feedback. 
Since the gas density at such small scales is $>10^{14}~\cc$ \citep{2012MNRAS.424..399G}, however,
the effect of dynamical friction would be underestimated.
Moreover, radiation from the formed binaries would determine their final properties as shown in \cite{Stacy_2016}.}.
The distribution of the initial separation is given by a Gaussian function with the average of 
$\langle a\rangle \simeq 0.4~{\rm AU}$ and the dispersion of $\sigma_a\simeq 0.34$ AU.
The primary mass and the mass ratio in this model have bimodal distributions, respectively
(see Table 2 of \citealt{2016arXiv161201524B} in detail).

\begin{figure}
\begin{center}
\includegraphics[width=83mm]
{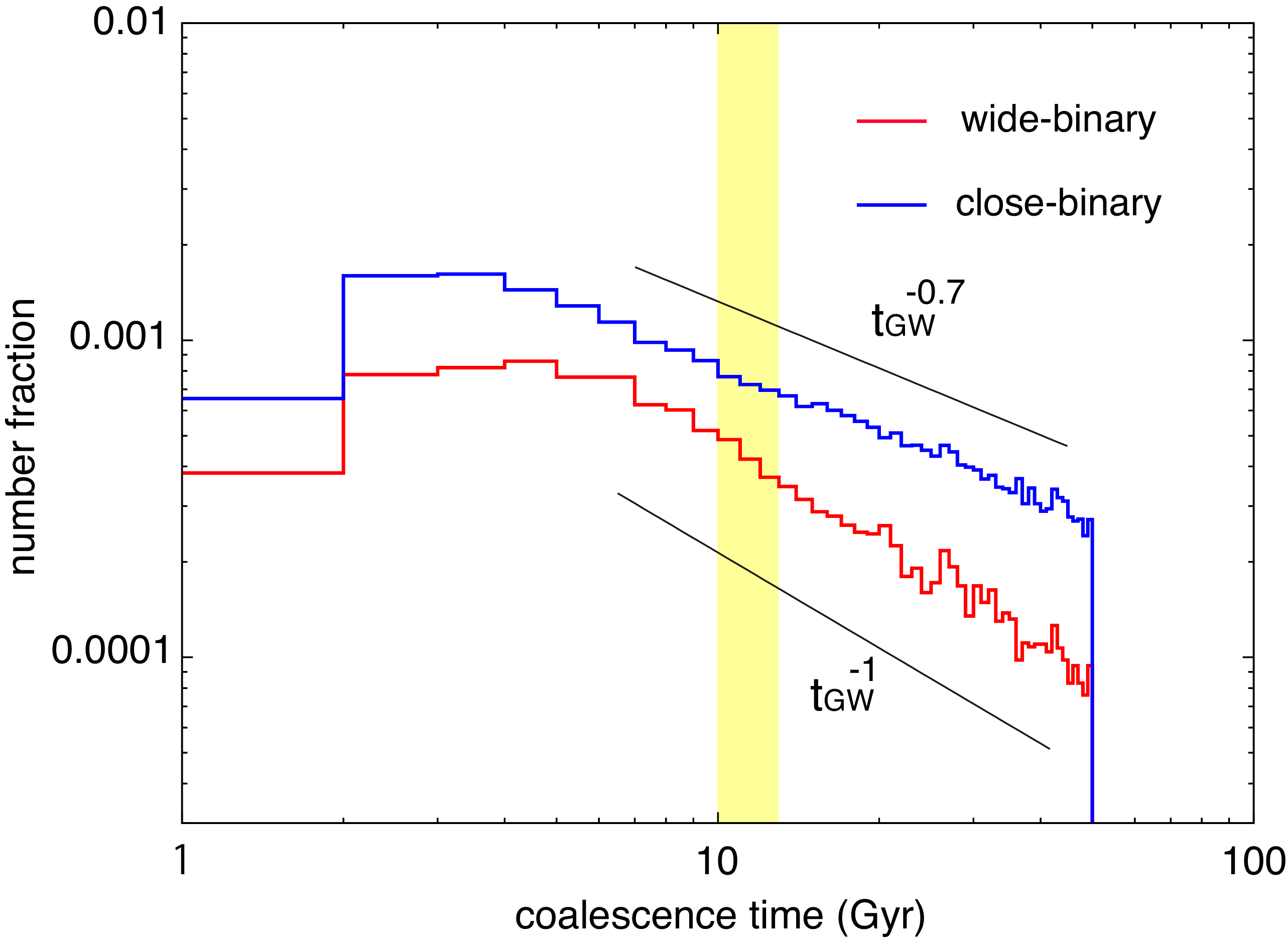}
\caption{Distribution of the coalescence time of PopIII BBHs 
for the wide-binary (red) and close-binary (blue) model, respectively.
Shaded region shows PopIII BBHs which can merge within the LIGO detection horizon.
}
\label{fig:tgw_dis}
\end{center}
\end{figure}

\subsection{Resultant distribution functions}
\label{sec:bdis}

Fig.~\ref{fig:Mch_hist} presents the chirp-mass distribution of PopIII BBHs for 
the wide-binary (red) and close-binary (blue) model, respectively.
Note that the chirp mass is a physical quantity to determine
the leading-order amplitude and frequency evolution of the GW signal.
The values of y-axis is normalized by the total number of PopIII binaries $N_{\rm tot}$;
PopIII BBHs formed through stable MT (dashed) and PopIII BBHs coalescing within the cosmic age, 
$t_{\rm GW}< 13.8$ Gyr (solid).
For the wide-binary model, $0.8\%$ of the PopIII binaries form BBHs through stable MT
and can merge due to GW emission within the Hubble time.
This result is consistent with the analytical estimate of the fraction shown in \S\ref{sec:analytic}, 
where $f_{\rm BBH}\simeq 2\%$ is estimated.
This difference is due to the simplification in Eq. (\ref{eq:f_BBH_4}).
In fact, PopIII BBHs with high chirp masses $35\la M_{\rm chirp}/\msun \la 45$ are formed, 
but do not merge because of their large separations (see also bottom panel in Fig.~\ref{fig:M1_q_a}).
On the other hand, for the close-binary model, 1.3\% of the PopIII binaries form BBHs 
through stable MT and and can merge due GW emission within the Hubble time.
In this model, the initial-separation distribution has a peak at $a_0\simeq 100~\rsun (\simeq 0.5$ AU)
and a tail toward smaller separations.
Thus, a significant fraction of PopIII binaries can form BBHs only via stable MT (see Fig.~\ref{fig:M1_q_a}).

Fig.~\ref{fig:tgw_dis} shows the distribution of the coalescence time of PopIII BBHs.
For the wide-binary model (red), the coalescence time distribution follows 
$dN/t_{\rm GW}\propto t_{\rm GW}^{-1}$
because of $\Gamma(a)\propto a^{-1}$.
For the close-binary model, since the initial separation distribution is approximated 
as $\Gamma(a)\propto a^\gamma$ with $\gamma \ga 0$ 
at $a \sim a_{\rm GW,0}$,
the coalescence time distribution is approximated as 
$dN/dt_{\rm GW}\propto t_{\rm GW}^{-0.7+(\gamma-0.2)/4}$.
Since most PopIII BBHs form at $z\simeq 10$, which corresponds to 
the cosmic age of $t_{\rm SF}\simeq 1$ Gyr, we require 
$10.8~{\rm Gyr} \la t_{\rm GW}+t_{\rm SF}\la 13.8~{\rm Gyr}$
for the PopIII BBHs to merge in the LIGO detection horizon 
($0\la z \la 0.2$, the corresponding time duration is $\sim 3$ Gyr).
From the distribution shown in Fig.~\ref{fig:tgw_dis},
the fraction of merging PopIII BBHs detectable by LIGO (i.e., $9.8 \la t_{\rm GW}/{\rm Gyr}\la 12.8$)
is estimated as $f_{\rm delay}\simeq 0.17$ for both models
and thus $f_{\rm BBH}f_{\rm delay}\simeq 1.7\times 10^{-3}$ is obtained.


\section{Observational constraints on PopIII BBH formation scenario}

In this section, we discuss the possibility that PopIII stars
can be the progenitors of the three BBH mergers detected by adLIGO, GW150914 \citep{Abbott_PRL_2016}, 
GW151226 and LVT151012 \citep{Abbott_PRL_2_2016} based on the formation efficiencies obtained in the
previous section and the total number of  PopIII stars inferred from Planck's result.

From the LIGO's detections, the stellar-mass BBH merger rate is estimated as $\mathcal{R}\simeq 55^{+99}_{-41}~\gpc^{-3}~\yr^{-1}$
 \citep{LIGO_BBH}.
Assuming that the BH mass function of coalescing binaries follows 
\begin{equation}
\Psi(M_1)\propto M_1^{-\alpha},
\end{equation}
with $M_{\rm min}\leq M_2 \leq M_1$ and $M_1+M_2\leq 100~\msun$,
and a uniform distribution on the secondary mass between $M_{\rm min}=5~\msun$ and $M_1$,
the power-law index is inferred as $\alpha \simeq 2.5^{+1.5}_{-1.6}$ \citep{LIGO_BBH}.

Let us suppose that all the LIGO events originate from PopIII BBHs.
We can evaluate the merger rate from the PopIII star formation rate  as 
\begin{equation}
\langle M_{\rm tot}\rangle \mathcal{R} \simeq 
\dot{\rho}_{\ast, \rm III}~\frac{2f_{\rm bin}}{1+f_{\rm bin}}~f_{\rm BBH}~f_{\rm delay},
\label{eq:merge}
\end{equation}
where $\dot{\rho}_{\ast, \rm III}$ is PopIII star formation rate (SFR), 
$f_{\rm bin}(\simeq 0.7)$ is the binary fraction,
$f_{\rm BBH}$ is the number fraction of merging BBHs within the Hubble time
to the total number of PopIII stars,
and $f_{\rm delay}$ is the number fraction of BBHs which merge in the LIGO detection horizon
to the total number of merging PopIII BBHs.
Here the mean mass of BH masses is $\langle M_{\rm tot} \rangle \simeq 30~\msun$ for $\alpha=2.5$. 
Therefore  the PopIII SFR required to explain all the LIGO events is 
\begin{align}
\dot{\rho}_{\ast, \rm III}&\simeq 1.3\times 10^{-3}~\msunyr \mpc^{-3}
\left(\frac{f_{\rm bin}}{1+f_{\rm bin}}\Big/ 0.4\right)^{-1}\nonumber\\
\times & \left(\frac{\langle M_{\rm tot}\rangle}{30~\msun}\right)
\left(\frac{\mathcal{R}}{60~\gpc^{-3}~\yr^{-1}}\right)
\left(\frac{f_{\rm BBH}}{0.01}\right)^{-1}
\left(\frac{f_{\rm delay}}{0.2}\right)^{-1},
\label{eq:SFR_1}
\end{align}
where we have used $f_{\rm BBH}$ obtained in \S\ref{sec:BBHeff}, i.e., 
the efficiency of BBH formation only via stable MT, which we consider as the lower limit.
For the two models for initial conditions of PopIII binaries, 
the number fraction of formed coalescing PopIII BBHs within the Hubble time is at least $1\%$,
i.e., $f_{\rm BBH}\ga 0.01$. In this sense, the estimated PopIII SFR can be lower by a factor of a few.
When we consider only GW\,150914, 
which has $\langle M_{\rm tot}\rangle \simeq 65.3~\msun$
and $\mathcal{R}=3.4^{+8.6}_{-2.8}~\gpc^{-3}~\yr^{-1}$,
the required PopIII SFR is $\dot{\rho}_{\ast, \rm III}\simeq 0.15\times 10^{-3}~\msunyr \mpc^{-3}$.

Now we compare directly the PopIII SFR of Eq.~(\ref{eq:SFR_1})
with the constraint by 
the Planck measurement of the optical depth of the universe to electron scattering, 
which arises due to ionizing photons emitted from PopIII stars and other sources
(e.g., accreting BHs).
Thus, assuming that all the photons required to explain the observed optical depth 
are produced only by PopIII stars,
the upper limit for the cumulative (comoving) mass density of PopIII stars is given by
\begin{align}
\rho_{\ast, \rm III}&\la 8.2\times 10^5~\msun~\mpc^{-3}\nonumber\\
&\times \left(\frac{\eta_{\rm ion}}{5\times 10^4}\right)^{-1}
\left(\frac{f_{\rm esc}}{0.1}\right)^{-1}
\left(\frac{\tau_{\rm e}-0.066}{\Delta \tau_{\rm e}}\right),
\label{eq:SFR_3}
\end{align}
\citep{Visbal_2015,Inayoshi_2016},
where $\eta_{\rm ion}$ is the number of H-ionizing photons per stellar baryon,
the escape fraction of ionizing photons from mini-haloes where PopIII stars form,
and $\tau_{\rm e}$ is the optical depth to electron scattering.
The ionizing photon number per baryon is $\eta_{\rm ion} = 7.1~(5.1)\times 10^4$ 
for the flat (Salpeter) IMF with a mass range of $10\leq M/\msun \leq 100$ \citep{2002A&A...382...28S}.
The value of the escape fraction $f_{\rm esc}\simeq 0.1$ corresponds to dark-matter halos with masses of 
$>{\rm a~few}\times 10^7~\msun$ \citep{Wise_2014}, where most of the PopIII stars would form 
(see also \citealt{Inayoshi_2016}).
We here set our fiducial value to $\tau_{\rm e}=0.066 \pm \Delta \tau_{\rm e}$,
where $\Delta \tau_{\rm e}=0.016$ is the 1$\sigma$ error \citep{Planck_2015}.
The PopIII SFR peaks at $z\simeq 10$ corresponding to the time scale of $\sim 0.5$~Gyr so that
the upper bound of PopIII SFR is 
\begin{equation}
\dot{\rho}_{\ast, \rm III} \la 1.3\times 10^{-3}~\msunyr \mpc^{-3}.
\end{equation}
This upper limit is consistent with the most probable value 
of the rate of all three events (see Eq.~\ref{eq:SFR_1}).
Therefore we conclude that the PopIII BBH formation scenario can explain
all the three BBH mergers detected in LIGO's O1 run 
with the maximal PopIII formation efficiency inferred from the Planck measurement,
even without BBHs formed by unstable MT or CE phases.


\section{Discussions}
\label{sec:discussion}

\subsection{Gravitational wave background}

A stochastic GW background (GWB) from unresolved PopIII BBH mergers
is a useful probe for the existence of massive BBH populations at higher redshifts
\citep{Hartwig_2016,Inayoshi_2016,Dvorkin_2016,Nakazato_2016}.
\citet{Inayoshi_2016} have estimated the amplitude of the PopIII GWB
adopting a PopIII star-formation rate whose normalization is consistent 
with the Planck measurement of the electron scattering optical depth
(see Eq \ref{eq:SFR_3}).
However, they also assume the merging rate of PopIII BBHs obtained by \cite{K14},
which include BBHs formed by unstable MT and CE phases.
According to the model by \cite{K14}, $12~(2.6)\%$ of PopIII binaries form 
BBHs coalescing within the Hubble time for the flat (Salpeter) IMF with $10\leq M/\msun \leq 100$
and $37~(55)\%$ of such BBHs (i.e., $4.2~(1.4)\%$ of the total binaries) are formed without experiencing CE phases.
On the other hand, the lower limit we estimate in \S\ref{sec:BBHeff} is only $\sim 1\%$ of total binaries.
Therefore, the GWB amplitude produced by merging PopIII BBHs formed only by stable MT
is several times smaller than that estimated by \cite{Inayoshi_2016}, but gives a lower limit of the GWB.
The detectability of the PopIII GWB in the future observing run in Advanced LIGO/Virgo in five years
depends on physical parameters relevant to reionization and high-redshift galaxies shown in Eq. (\ref{eq:SFR_3}).

\subsection{Tidal force and stellar spins}
\label{sec:spin}

In this paper, we neglect effects of stellar rotation on the evolution so far.
As we discussed, non-rotating PopIII stars are likely to evolve to compact blue giants 
and thus can avoid unstable MT and CE phases during their lifetime.
However, stellar rotation may change the evolution of Pop III stars
because of hydrodynamical instabilities induced by rotation (e.g. meridional circulations).
For a slow-rotating PopIII star with $v_{\rm rot}\sim 0.2~v_{\rm Kep}$ 
($v_{\rm Kep}$ is the Keplerian velocity of the star), the effective temperature at the end of the 
main-sequence phase tends to be lower than for a non-rotating PopIII star with the same mass,
because unburnt hydrogen is supplied to the core of the rotating PopIII star due to the mixing
\citep{Ekstrom_2008,Takahashi_2014}.
Furthermore, mixing of heavy elements would increase the opacity of the stellar envelope
\citep{Joggerst_2011}.
Therefore, the rotating PopIII star tends to evolve to red giants instead of blue giants expected 
for the non-rotation case.
Since a red giant has a bloated stellar convective envelope, 
the stellar radius is more likely to fill its Roche lobe and lead to unstable MT,
which would result in CE phase.
On the other hand, a fast-rotating PopIII star with $v_{\rm rot}\ga 0.5~v_{\rm Kep}$,
would experience chemically homogeneous evolution due to strong mixing effects and 
then become bluer without any redwards evolution \citep[e.g.][]{Yoon_2005,Woosley_Heger_2006,
deMink_2009,Song_2016}.
In this case, the PopIII stars can avoid unstable MT and CE phases.

The rotation period of a star in a close binary system is likely to be the same as 
the orbital period because of the tidal torque. 
This synchronization timescale of massive main sequence stars is 
estimated as (\citealt{Kushnir_2016a,Kushnir_2016b}; see also \citealt{Zhan_1975}):
\begin{align}
t_{\rm tide} &\simeq  2~{\rm Myr}~
q^{-1/8}\left(\frac{1+q}{2q}\right)^{31/24}
\left(\frac{r_{\rm g}^2}{0.1}\right)
\left(\frac{t_{\rm GW}}{10~{\rm Gyr}}\right)^{17/8}\nonumber\\
\times &\left(\frac{R}{3~\rsun}\right)^2
\left(\frac{R_{\rm con}}{1.2~\rsun}\right)^{-9}
\left(\frac{M}{40~\msun}\right)^{109/24}
\left(\frac{M_{\rm con}}{24~\msun}\right)^{4/3},
\end{align}
where $R_{\rm con}$ and $M_{\rm con}$ are the size and mass of the convective core,
$r_{\rm g}^2$ is the the gyration radius of the star.
Here the fiducial values are adopted from data calculated with the {\tt MESA} code 
and approximate $\rho_{\rm con}/\bar{\rho}_{\rm con}\simeq 0.5$. 
PopIII main sequence stars in binaries are likely to be synchronized at
orbital separations where the coalescence time is shorter than the Hubble time. 
Note that this estimate is very sensitive to the convective-core radius and the 
mass of the star.

Importantly, when they are tidally synchronized at the separation
where the coalescence time corresponds to the Hubble time,
the rotational velocity of synchronized PopIII main sequence stars
 is quite slow as $v_{\rm rot} < (0.01-0.1) \times v_{\rm Kep}$. 
Therefore, the stellar rotation does not play important roles irrespective with
the initial spin. Of course, binaries with smaller separations, i.e., shorter coalescence times,
the stellar rotation might play important roles. However, BBHs formed in such systems 
merge in the early universe and they are not observable by LIGO.

\subsection{Remnants after common envelope phases}
\label{sec:CE}

Throughout the paper, we do not discuss the outcome of PopIII binaries which experience CE phases,
in order to give conservative arguments.
Many previous studies concluded that possible outcomes after CE evolution are tight binaries 
and/or stellar mergers, but the bifurcation conditions are highly uncertain 
\citep[e.g.,][]{Iben_1993,Taam_2000,Ivanova_2013}.
Even for PopIII binaries, CE evolution would occur for massive stars with $M > 50~\msun$
because (1) the accretor expands due to violent MT from the donor or 
(2) the MT would be unstable when the donor has a deep convective envelope 
in the late stage of its evolution (see bottom panel in Fig.~\ref{fig:M1_q_a}).
If the CE evolution of such binaries does not lead to stellar mergers,
possible types of the remnants would be binaries of a He star ($\ga 20~\msun$) 
with a main-sequence star ($\ga 40~\msun$) or a BH ($\ga 30~\msun$)
after the first/second episodes of MT.
Subsequently, if any, they form massive BBHs coalescing due to GW emission 
within the Hubble time.
In order to discuss the outcome after CE evolution of PopIII binaries, 
we need to understand basic properties of the stellar dynamics in the CE phases, 
e.g., the energy budget during the CE phases more precisely using stellar evolution calculations
as studied for stars with $0.0004 \leq Z \leq 0.02$ \citep[e.g.,][]{2016A&A...596A..58K}.

\section{Summary}
\label{sec:summary}
We study formation of stellar mass BBHs originating from PopIII stars, 
performing stellar evolution simulations for PopIII binaries with a public code {\tt MESA}. 
We find that a significant fraction of PopIII binaries form massive BBHs 
through stable MT without experiencing CE phases. 
The formation efficiency of coalescing PopIII BBHs is estimated for two different 
initial conditions for PopIII binaries with large and small separations, respectively. 
As a result, $\sim 10 \%$ of the total PopIII binaries form BBHs only through stable MT
and $\sim 10\%$ of these BBHs merge due to gravitational wave emission within the Hubble time. 
Furthermore, the chirp mass of merging BBHs has a flat distribution over $15\la M_{\rm chirp}/\msun \la 35$. 
This formation pathway of PopIII BBHs is presumably robust because stable MT
is less uncertain than CE evolution, which is the main formation channel 
for PopII BBHs.  
We then test the hypothesis that the BBH mergers detected
by LIGO originate from PopIII stars using our result and the upper limit on the 
total number of PopIII stars formed
in the early universe as inferred from the optical depth measured by Planck.
We conclude that the PopIII BBH formation scenario can explain
the mass-weighted merger rate of the LIGO's O1 events
with the maximal PopIII formation efficiency inferred from the Planck measurement,
even without PopIII BBHs formed by unstable MT or CE phases.

\section*{Acknowledgements}
We thank Georges Meynet for improving the paper as a referee.
We also thank Zolt\'an Haiman, Takashi Hosokawa, Sylvia Ekstr$\ddot{\rm o}$m and 
Tilman Hartwig for useful discussions.
This work is partially supported by the Simons Foundation through 
the Simons Society of Fellows (KI), Flatiron Fellowship (KH), 
by JSPS Research fellowship for young scientists (RH, No. 16J07613), and
by JSPS KAKENHI Grant (TK, No. JP16818962).

\bibliographystyle{mnras}
{\small
\bibliography{ref}
}


\appendix

\section{Non-conservative mass transfer: the dependence on $\beta$}

Through the paper, we set the value of $\beta$, which is a parameter to 
describe the efficiency of non-conservative mass transfer.
We adopt $\beta=0.1$ as our fiducial value.
In this appendix, we briefly discuss the dependence of the choice of $\beta$ 
on the results.
The orbital evolution is described by Eq. (\ref{eq:a_1}).
Integrating this equation over the second episode of MT, where 
the donor is a star with $M_1$ and the accretor is a BH with $M_2$,
the orbital separation after the MT is given by
\begin{equation}
a_f=
\left(\frac{M_{1i}}{M_{1f}}\right)^2
\left(\frac{M_{2i}}{M_{2f}}\right)^{2/\beta}
\left(\frac{M_{1i}+M_{2i}}{M_{1f}+M_{2f}}\right)a_i,
\end{equation}
where the subscript $i$ and $f$ indicate the values before and after the MT, respectively,
and $M_{2f}=M_{2i}+\beta(M_{1i}-M_{1f})$.
For $\beta=0$, $M_2$ does not change via the MT, we obtain 
\begin{equation}
a_f=
\left(\frac{M_{1i}}{M_{1f}}\right)^2
\left(\frac{M_{1i}+M_{2}}{M_{1f}+M_{2}}\right)
e^{2(q_{1f}-q_{1i})}a_i.
\label{eq:app1}
\end{equation}

\begin{figure}
\begin{center}
\includegraphics[width=83mm]
{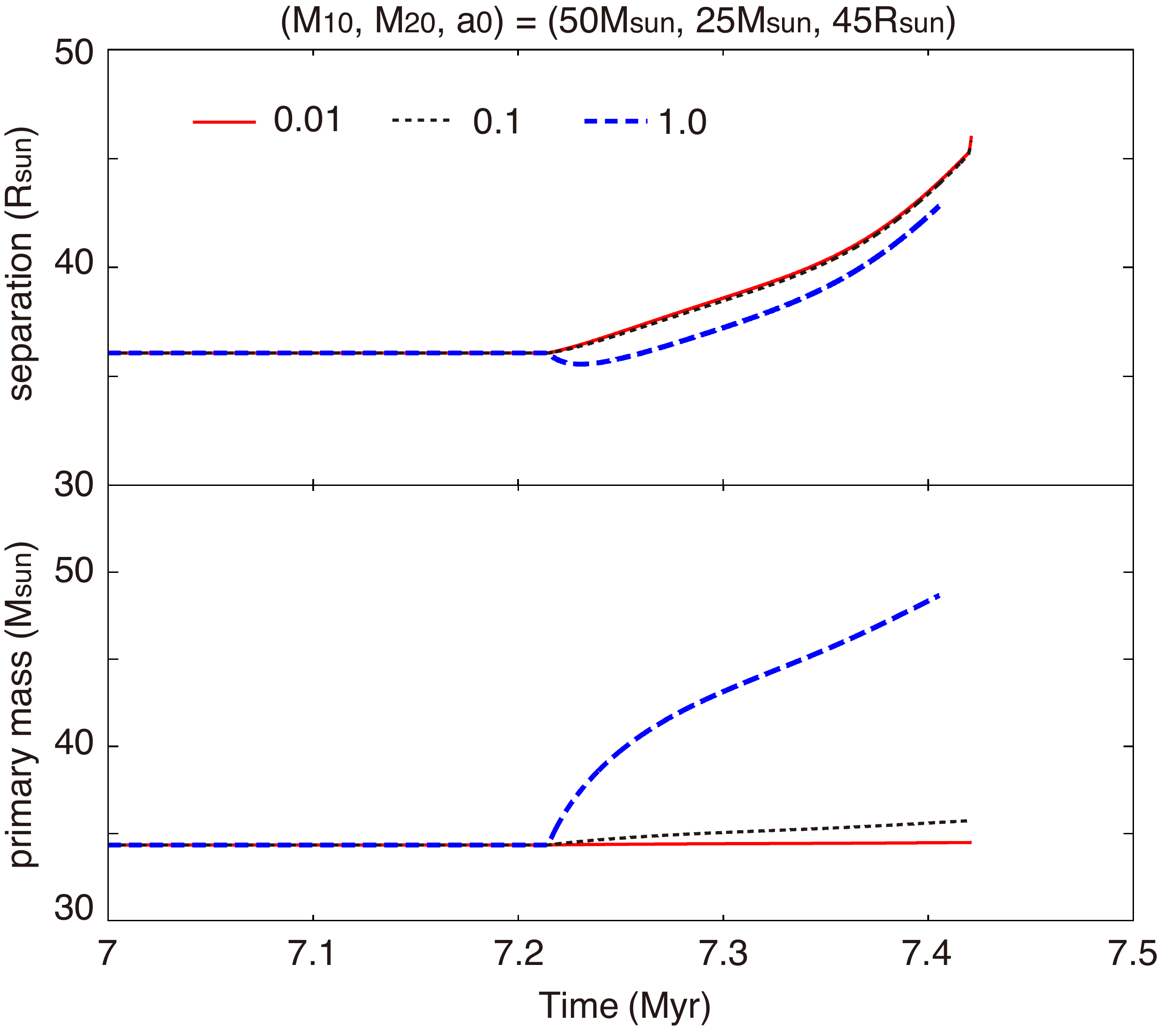}
\caption{Time evolution of the orbital separation and the BH (accretor)
with initial conditions of $M_{1,0}=50~\msun$, $M_{2,0}=25~\msun$ and $a_0=45~\rsun$
for different values of $\beta$, which describes the efficiency of non-conservative mass transfer.
}
\label{fig:beta_dep}
\end{center}
\end{figure}

Fig. \ref{fig:beta_dep} presents the time evolution of the orbital separation 
and the BH mass with the same initial conditions as shown in 
Figs. \ref{fig:50-25M_45R} and \ref{fig:semi_50_25-45}
for different values of $\beta$.
For $\beta \la 0.1$, the resultant track of the separation and the BH (accretor) 
mass do not change significantly since the final separation can be written by
Eq. (\ref{eq:app1}) for smaller $\beta$.
For conservative MT ($\beta=1$), the final orbital separation is similar to 
those for $\beta \la 0.1$, but the BH mass becomes higher.
However, a significant fraction of the transferred mass would be ejected 
from the system 
because the accretion flow onto the BH releases
a huge amount of energy as radiation and/or outflows
\citep[e.g.][]{Blandford_Begelman_1999, 2005ApJ...628..368O,2014ApJ...796..106J,2015MNRAS.447...49S}.

\end{document}